\documentclass[a4paper,fleqn,usenatbib,useAMS]{mnras}

\usepackage{graphicx}   % Including figure files
\usepackage{amsmath}    % Advanced maths commands
\usepackage{amssymb}    % Extra maths symbols
\usepackage{subfigure}
\usepackage{epstopdf}

\newtheorem{theorem}{Theorem}[section]

\usepackage[T1]{fontenc}
\usepackage{ae,aecompl}

\usepackage{txfonts}
\usepackage{mathptmx}
\title[Perturbative solution to the Lane-Emden equation]{Perturbative solution to the Lane-Emden equation: An eigenvalue approach}
\author[Kenny L. S. Yip, T. K. Chan, P. T. Leung]{Kenny L. S. Yip$^{1}$\thanks{Email address: \href{mailto:yiplongsang@gmail.com}{yiplongsang@gmail.com}},
T. K. Chan$^{1,2}$\thanks{Email address:
\href{mailto:tkc004@physics.ucsd.edu}{tkc004@physics.ucsd.edu}},
P. T. Leung$^{1}$\thanks{Email address:
\href{mailto:ptleung@phy.cuhk.edu.hk}{ptleung@phy.cuhk.edu.hk}}
\\
$^{1}$Department of Physics, the Chinese University of Hong Kong, Shatin, N.T., Hong Kong \\
$^{2}$Department of Physics, University of California at San Diego, 9500 Gilman Drive, La Jolla, CA 92093, USA}

\date{\today}

\pubyear{2016}

\begin{document}
\label{firstpage}
\pagerange{\pageref{firstpage}--\pageref{lastpage}}
\maketitle

\begin{abstract}
Under suitable scaling, the structure of self-gravitating
polytropes is described by the standard Lane-Emden equation (LEE),
which is characterised by the polytropic index $n$. Here we
use the known exact solutions of the LEE at $n=0$ and $1$
to solve the equation perturbatively. We
first introduce a scaled LEE (SLEE) where polytropes with
different polytropic indices all share a common scaled radius. The
SLEE is then solved perturbatively as an eigenvalue problem.
Analytical approximants of the polytrope function, the radius and
the mass of polytropes as a function of $n$ are derived. The
approximant of the polytrope function is well-defined and
uniformly accurate from the origin down to the surface of a
polytrope. The percentage errors of the radius and the mass are
bounded by $8.1 \times 10^{-7}$ per cent and $8.5 \times 10^{-5}$
per cent, respectively, for $n\in[0,1]$. Even for $n\in[1,5)$,
both percentage errors are still less than $2$ per cent.
\end{abstract}


%%%%%%%%%%%%%%%%%%%%%%%%%%%%%%%%%%%%%%%%%%%%%%%%%%

%%%%%%%%%%%%%%%%% BODY OF PAPER %%%%%%%%%%%%%%%%%%

\section{Introduction} \label{sec:introduction}
The structure and the dynamics of stars and collisionless galaxies
(or clusters) are often  mimicked by polytropes  characterised by
a polytropic index $n$ \citep[see,
e.g.,][]{chandrasekhar1958introduction,Cox,binney}. For example,
stars with a vanishing polytropic index are incompressible,
whereas stars with $n \ge 3$ are unstable against radial
oscillations. In general, the stiffness of a polytropic star
decreases with an increase in the polytropic index. On the other
hand, the spatial extent of a polytropic galaxy (or cluster) is
infinite for $n \ge 5$, and in the isothermal sphere model of
galaxies the $n \rightarrow \infty $ limit of polytropes is indeed
considered. The physics of a self-gravitating polytrope in the
Newtonian limit is governed by the famous Lane-Emden equation
(LEE) \citep[see, e.g.,][Section 3.3]{Cox}, which is non-linear in
the so-called polytrope function, i.e., the solution of the LEE
(see Section \ref{sec:homology_invariance} for the exact form of
the LEE and the definition  of the polytrope function). Physically
speaking, the polytrope function $\theta$ itself is related to the
gravitational potential and $\theta^n$ directly measures the mass
density of a polytrope.

In addition to its application in astrophysics as mentioned above,
the LEE is also interesting in its own right \citep[see,
e.g.,][and references therein]{Ramos2008400}. Except for some
special values of the polytropic index, namely $n=0,1$ and $5$,
the LEE does not have any known analytical solutions which are
expressible in terms of elementary functions. Series solutions for
the LEE have been sought and the recursion relation of such a series
was derived \citep{Seidov_Series}. However, the series solution to
the LEE about the center of a polytrope diverges before reaching
the surface of the polytrope if $n>1.9121$ \citep{Hunter}. On the
other hand, even for $0<n<1.9121$, the convergence rate of a
formal series solution of the LEE could be slow and is in general
dependent on both $n$ and the position variable. In order to
remedy the problem of divergence or the slow convergence of the series
solution, methods of Pad\'{e} resummation and variable
transformation have been employed to extend the interval of
validity and accelerate the convergence of the series
\citep{Pascual_Pade, Iacono,Ramos2008400}.  For example,  in order
to extend the radius of convergence of the series solution,
\citet{Roxburgh} used the polytropic mass as the independent
variable, in place of the radial distance, to  extend the interval
of convergence down to the surface of polytropes but thousands of terms
are needed in order to achieve satisfactory accuracy near the
stellar surface.

Instead of pursuing exact analytical solutions of the LEE, a lot
of approximation schemes have also been proposed. Some interpolated
from exact analytical solutions and/or numerical results.
Given the analytically
closed form solutions to the LEE at $n=0, 1$ and $5$,
\citet{Buchdahl_n5} proposed a rational function containing three
free parameters to match the analytical solution at these special
values of $n$. Following Buchdahl, \citet{Iacono} derived an
analytical approximant of the radius but using numerical data
resulting from Runge-Kutta integrations. \citet{Liu_fit} solved
the LEE approximately in various parameter and spatial regimes and
combined these solutions together with some empirical fitting
constants to yield accurate approximate solutions of the equation.

The LEE is a nonlinear differential equation except for $n=0$ and
$1$. To generate analytical approximate solutions to the LEE,
\citet{Bender_delta_expansion} applied the delta expansion method
(DEM),  which was proposed to solve a wide range of non-linear
problems through expanding the non-linear term in a power series
of its exponent, and succeeded in expanding the polytrope function
into a power series of $n-1$ about the exact solution at $n=1$.
As LEE admits a closed form solution at $n=0$ as well,
\citet{Seidov_LEE} considered the DEM about the incompressible
limit where $n=0$, which leads to the expansion of the polytrope
function into powers of $n$. However, both of these two attempts
encountered a common problem, namely the expansion of the
perturbed polytrope function becomes singular at the surface of
the unperturbed polytrope where the unperturbed polytrope function
vanishes. As a result, the perturbed polytrope function might
become complex-valued near the surface of the perturbed polytrope.
%Here, we take into account of the $n$-dependence of the
%singularity by means of a $n$-dependent transformation of the
%length scale that polytropes share a common radius at $\pi$ in the
%new scale.

The goal of the present paper is to deduce analytical
approximations of various physical quantities of polytropes,
including the polytrope function, the radius and the mass, as a
function of the polytropic index. In particular, we adopt the DEM
proposed by \citet{Bender_delta_expansion} and \citet{Seidov_LEE}
to handle the nonlinear nature of the LEE.  In order to remedy the
singular behaviour of the polytrope function outside the
unperturbed polytrope, we propose to scale the radius of
polytropes in such a way that both the perturbed and unperturbed
polytropes share a common radius of $\pi$ in the new length scale.
Hence, the perturbation solution of the polytrope function is
analytic inside the entire perturbed polytrope. As a result, our
method, referred to as the scaled delta expansion method (SDEM) in
the present paper, is able to yield accurate approximants of the
polytrope function, the radius and the mass of a polytrope.  To
our knowledge, it is the first successful perturbation attempt
that properly takes into account of the branch point singularity
on the surface of a polytrope, which is essential in analytical
evaluation of various physical quantities, such as the radius, the
mass, the moment of inertia, the tidal deformability, to name a
few. In the present paper, we restrict our attention to finite
polytropes (i.e., polytropes of finite radius), where the
polytropic index ranges from $0$ inclusively to $5$ exclusively.

Applying the SDEM in tandem with the Pad\'{e} approximation method
\citep[see, e.g.,][]{baker1975essentials},  we derive analytical
global approximants of the radius and the mass of polytropes as a
function of the polytropic index $n$, which respectively have the
maximum percentage errors of $8.1 \times 10^{-7}$ per cent and
$8.5 \times 10^{-5}$ per cent for $n$ in $[0,1]$. In general, for
$n$ lying in $[0,5)$, where polytropes are of finite spatial
extents, the said errors are still less than $1$ per cent and $2$
per cent, respectively.

The organization of this article is outlined below.  In Section
\ref{sec:homology_invariance}, we derive the LEE from the
assumptions of hydrostatic equilibrium and the polytropic equation
of state, with emphasis on the relationship between the freedom of
the choice of the density scale and the scale invariance symmetry
of the LEE. Next, in Section \ref{sec:literature_review1}, we
review the applications of the DEM, as proposed by
\citet{Bender_delta_expansion} and \citet{Seidov_LEE}, to the LEE.
In Section \ref{sec:homology_perturbation}, we introduce the SDEM.
We exploit the scale invariance symmetry of the LEE to define a
length scale depending on the polytropic index, such that we take
into account of the moving singularity due to the first zero of
the solution (i.e., the polytrope function) of the LEE by a
variable scale. As such, the size of polytropes in the transformed
length scale remains the same throughout the course of
perturbation analysis. Using the SDEM, we find analytical
approximations of the polytrope function and  the radius of
polytropes as a function of the polytropic index, and compare our
numerical results
 with other
existing approximations. In Section \ref{sec:approximants_mass},
we further apply the SDEM to derive analytical approximations of
the mass of a polytrope as a function of the polytropic index $n$
about each perturbation center of the SDEM. In Section
\ref{sec:global_approximants} we interpolate the local
approximants developed about different perturbation centers
through the Pad\'{e} approximation method to yield uniform
approximants of the radius, the mass and the polytrope function
of polytropes with high accuracy for $0\le n<5$. We then conclude
our paper in Section \ref{sec:conclusion}. In Appendix
\ref{sec:literature_review23}, for ease of reference, we provide
the results of some previous attempts of approximating the
solution to the LEE by series expansion, interpolation and
fittings. In Appendix
\ref{sec:formulae_for_lazy_physicists}, we list some useful
formulae derived here  in a self-contained manner, so that
astrophysicists interested in the solution of the LEE could apply
them readily.

\section{The Lane-Emden Equation} \label{sec:homology_invariance}

The LEE is conventionally expressed as:
\begin{equation}
\mathcal{L}_{x} \theta(x) + [\theta(x)]^{n} = 0, \label{eqn:LEE}
\end{equation}
where the solution $\theta(x)$ is called  the polytrope function
and the operator $\mathcal{L}_{s}$ is  defined as:
\begin{equation}
\mathcal{L}_{s} \equiv \frac{1}{s^2} \frac{d}{ds} \left( s^2
\frac{d} {ds} \right)~.
\end{equation}
Physically speaking, the LEE directly follows from the Poisson
equation of Newtonian gravity, and the equilibrium condition
of a self-gravitating   polytropic star (or a collisionless
galaxy/cluster):
\begin{equation}
\label{eqn:hydrostaticequilibrium} \frac{1}{r^2} \frac{d}{dr}
\left[ \frac{r^2}{\rho(r)} \frac{d P(r)}{dr} \right] = - 4 \pi G
\rho(r),
\end{equation}
where $\rho(r)$ and $P(r)$ are respectively the mass density and
the pressure at a radius $r$, $G$ is the constant of universal
gravitation.
For a specific polytropic equation of state:
\begin{equation}
\label{eqn:polytropic}
P(r) = K \rho(r)^{1 + 1/n},
\end{equation}
where $K >0$ and the polytropic index $n \geq 0$ are given
parameters,  one can introduce an \emph{arbitrary} density scale
$\rho_0>0$ and an associated length scale $a$:
\begin{equation}
\label{eqn:lengthscale} a = \sqrt{\frac{K (n+1)}{4 \pi G}}
\rho_{0}^{(1-n)/(2n)},
\end{equation}
to define a dimensionless radius $x = r/a$ and the polytrope
function $\theta(x)$:
\begin{equation}
[\theta(x)]^{n} \equiv \frac{\rho(r)}{\rho_{0}} \label{eqn:density_definition}  .
\end{equation}
It is then straightforward to show that the polytrope function
$\theta(x)$ satisfies the LEE \eqref{eqn:LEE}. Physically
speaking, $[\theta(x)]^{n}$ is a measure of the density
distribution. Besides, it is readily shown that $\theta(x)$ is, up
to an additive constant, proportional to the gravitational
potential.

It is worthwhile to note that the freedom of the choice of the
density scale $\rho_{0}$, hereafter referred to as the scale
invariance symmetry of the LEE, also leads to the freedom of the
initial condition for $\theta(x)$ and the length scale $a$.    We
shall see that such a symmetry motivates the SDEM in Section
\ref{sec:homology_perturbation}. On the other hand, it is
customary to choose the following initial conditions of the LEE:
\begin{equation}
%\begin{split}
\hat{\theta}(0)=1, \quad \frac{d \hat{\theta}}{dx}(0)=0.
%\end{split}
\label{eqn:IC9}
\end{equation}
Hereafter we use $\hat{\theta}(x)$ to indicate normalised
polytrope functions satisfying the above initial conditions. While
the first condition $\hat{\theta}(0)=1$ is equivalent to the
assumption that $\rho_{0}$ equals the central density $\rho_c
=\rho(r=0)$, the second one is indeed a direct consequence of the
boundedness of $\rho_c$. It should be noted that, in general, we
may choose any $\theta(0)>0$.

It is well known that the LEE admits closed form solutions for
$n=0,1$ and $5$ as follows:
\begin{align}
& n=0,\quad \hat{\theta}(x) = 1- \frac{1}{6} x^{2}, \quad \hat{\xi} = \sqrt{6}, \label{eqn:exact0} \\
& n=1, \quad \hat{\theta}(x) = \frac{\sin(x)}{x}, \quad \hat{\xi} = \pi, \label{eqn:exact1}\\
& n=5, \quad \hat{\theta}(x) = \frac{1}{\sqrt{1 + \frac{x^{2}}{3}}}, \quad \hat{\xi} = \infty, \label{eqn:exact5}
\end{align}
where $\hat{\xi}$ is the first zero of the normalised polytrope
function $\hat{\theta}(x)$, and $\hat{\xi} = \infty$ means the
solution does not vanish on the positive real line.
These solutions could be easily verified
by direct substitution \citep[see, e.g.,][equations (3) -
(5)]{Seidov_LEE} and used as the starting point of perturbation
analysis as well as good check of numerical calculations.

In general, the first zero of $\theta(x)$, $\xi$, depends on the
density scale $\rho_0$. $\xi$ is of physical interest because it
is related to the physical radius $R$ of a polytropic star by:
\begin{equation}
\label{eqn:polytrope_radius}
R  = a \xi = \sqrt{\frac{K (n+1)}{4 \pi G}} \rho_{c}^{(1-n)/(2n)} \left[ \theta(0)^{(n-1)/2} \xi \right]. \\
\end{equation}
As the first zero of a polytrope function $\xi$ is the dimensionless counterpart
of the physical radius of a polytrope, we will use the terms
first zero and radius interchangeably if no ambiguity arises.
By the same token, the first zero of a normalised polytrope function
will be referred to as the normalised first zero and
the normalised radius interchangeably.
We will emphasise $R$ as the \emph{physical} radius.
We see that the consequence of the scale invariance of the LEE
extends to all physical quantities of polytropes.  Any physical
quantities of a polytrope are uniquely determined by its central
density $\rho_c$, the parameter $K$ and the polytropic index $n$,
and do not depend on our particular choice of $\theta(0)$.
Therefore, the combination $\theta(0)^{(n-1)/2} \xi$ appearing in
equation \eqref{eqn:polytrope_radius} must remain the same
regardless of our choice of the initial condition $\theta(0)$.  In
other words, $\theta(0)^{(n-1)/2} \xi$ is a scale-invariant
combination signifying the physical radius of a polytropic star (or
galaxy/cluster). In this regard, the normalised polytrope function
$\hat{\theta}(x)$ and its associated radius $\hat{\xi}$ are
not particularly superior to other solutions to the LEE. In the
following discussion, we shall make use of such a symmetry to remedy
the singularity problem encountered in the DEM considered by
\citet{Bender_delta_expansion} and \citet{Seidov_LEE}.

\section{Delta expansion method} \label{sec:literature_review1}

Since closed form solutions to the LEE are only known for $n=0,1$
and $5$, many articles have been devoted to the analytical
approximations of the LEE at values of $n$ other than $0,1$ and
$5$.  In particular, the physical radius of a polytrope is given by the
first zero of the associated polytrope function as shown in
equation (\ref{eqn:polytrope_radius}).  It is of physical interest
to determine the first zero as a function of the polytropic index
$n$.  A lot of studies on analytical approximants of the LEE have
been performed  using various techniques, including series
expansion methods \citep[see,
e.g.,][]{Seidov_Series,Hunter,Pascual_Pade, Iacono}, perturbation
methods \citep[see, e.g.,][]{Bender_delta_expansion,Seidov_LEE},
and empirical interpolation schemes \citep[see, e.g.,][]{Liu_fit}.
Here, we review in depth the DEM of the LEE
\citep{Bender_delta_expansion,Seidov_LEE}, because it is one of
the foundations of the SDEM, which we are going to propose in
Section \ref{sec:homology_perturbation}. For ease of reference, a
brief summary of the results obtained in other previous studies on
the LEE is also provided in Appendix
\ref{sec:literature_review23}.

\citet{Bender_delta_expansion} first introduced the DEM to solve a
wide range of non-linear problems by considering the exponent of a
non-linear term as the perturbation parameter. As a result, the
non-linear term is expanded into a power series of the exponent. In
particular, \citet{Bender_delta_expansion} applied the DEM to the LEE
by expanding $[\hat{\theta}(x)]^{n}$ in equation \eqref{eqn:LEE}
into a power series of $n-1$ about the exact solution at $n=1$. As
LEE admits closed form solutions at $n=0$ as well,
\citet{Seidov_LEE} considered the DEM about $n=0$ and the
$[\hat{\theta}(x)]^{n}$ term is expressed in terms of a power
series in $n$. With such an expansion, \citet{moment} studied the
relationship among the multipole moments of compact stars in the
Newtonian regime.

For illustration, we outline the application of the DEM to the LEE
about the incompressible limit where $n=0$ \citep{Seidov_LEE}.
Under the assumption that the polytropic index $n$ is small, the
normalised polytrope function $\hat{\theta}(x) =
\hat{\theta}_{\text{S}}(x)$ and the non-linear term
$[\hat{\theta}_S(x)]^{n}$ are  expanded into their respective
power series in $n$ as follows:
\begin{align}
& \hat{\theta}_{\text{S}}(x) = \hat{\theta}_{\text{S}}^{(0)}(x) + n
\hat{\theta}_{\text{S}}^{(1)}(x) + n^2 \hat{\theta}_{\text{S}}^{(2)}(x) +
O[n^3], \\
& \hat{\theta}_{\text{S}}^{n} = 1 + n \ln \hat{\theta}_{\text{S}}^{(0)} + n^2
\left( \frac{ \hat{\theta}_{\text{S}}^{(1)}}{ \hat{\theta}_{\text{S}}^{(0)}} +
\frac{1}{2} \ln^2 \hat{\theta}_{\text{S}}^{(0)} \right) + O[n^3].
\end{align}
  As a result, the LEE is
reduced to a system of coupled differential equations.  The
remainder of the problem is to solve these equations to find
$\hat{\theta}_{\text{S}}^{(0)}$, $\hat{\theta}_{\text{S}}^{(1)}$ and
$\hat{\theta}_{\text{S}}^{(2)}$ recursively. Up to the second order, the
 system of differential equations reads:
\begin{align}
&{\cal L}_x \hat{\theta}_{\text{S}}^{(0)}= -1,\\
&{\cal L}_x \hat{\theta}_{\text{S}}^{(1)} = -\ln \hat{\theta}_{\text{S}}^{(0)}
,\\
&{\cal L}_x \hat{\theta}_{\text{S}}^{(2)}
= - \frac{ \hat{\theta}_{\text{S}}^{(1)}}{ \hat{\theta}_{\text{S}}^{(0)}} -
\frac{1}{2} \ln^2 \hat{\theta}_{\text{S}}^{(0)},
\end{align}
which are subject to the initial conditions:
\begin{equation}
\begin{split}
\hat{\theta}_{\text{S}}^{(i)}(0)=\delta_{0i}, \quad \frac{d
\hat{\theta}_{\text{S}}^{(i)}}{dx}(0) = 0,
\end{split}
\end{equation}
for $i=0,1,2,\ldots$.

\citet{Seidov_LEE} gave the perturbation solution to the polytrope
function, $\hat{\theta}_{\text{S}}(x)$ about $n=0$ up to the first
order in $n$:
\begin{align}
\nonumber \hat{\theta}_{\text{S}}(x) = &  1-\frac{x^2}{6}+n \bigg[\frac{5
x^2}{18}-4+
\left(3-\frac{2 \sqrt{6}}{x}-\frac{x^2}{6}\right) \ln \left(1-\frac{x}{\sqrt{6}}\right) \\
& + \left(3+\frac{2\sqrt{6}}{x}-\frac{x^2}{6}\right) \ln
\left(1+\frac{x}{\sqrt{6}}\right)\bigg] +
O[n^2],\label{eqn:Seidov_theta}
\end{align}
and the normalised radius,  $\hat{\xi}_{\text{S}}(n)$, as a function of
$n$ up to the second order in $n$:
\begin{align}
\nonumber \hat{\xi}_{\text{S}}(n) & = \sqrt{6}+ n \frac{ \sqrt{6} }{6} \bigg( -7+12 \ln 2 \bigg) + n^2 \frac{\sqrt{6}}{72} \bigg(1379-84 \pi ^2 \\
\nonumber & \quad -888 \ln 2+144 \ln^2 2\bigg) + O[n^3] \\
& \approx \sqrt{6}+0.537975784794 n+0.123283090086 n^2 +
O[n^3].\label{eqn:Seidov_xi}
\end{align}
The algebraic details could be found in his work
\citep{Seidov_LEE}. Notice that the logarithmic term
$\ln(1-{x}/\sqrt{6})$ in the expression of
$\hat{\theta}_{\text{S}}(x)$ becomes complex when $x
> \sqrt{6}$. However, as shown in equation \eqref{eqn:Seidov_xi}, the interval of physical
interest of $x \in [0 , \hat{\xi}_{\text{S}}(n)] $ extends beyond
$\sqrt{6}$ for $n
> 0$. To get a real solution of $\hat{\xi}_{\text{S}}(n)$ by method of analytic continuation, in the following
we will take the real part of the  expansion in equation
\eqref{eqn:Seidov_xi} for $x > \sqrt{6}$.

By the same token, \citet{Bender_delta_expansion} derived the
perturbation solution to the polytrope function,
$\hat{\theta}_{\text{B}}(x)$, about $n=1$ to the first order in
$n-1$,
\begin{align}
\nonumber \hat{\theta}_{\text{B}}(x) = & \frac{\sin x}{x} + (n-1) \bigg[ \frac{\cos x }{2 x} \int_{0}^{x} \ln \sin t dt + \frac{3}{4} \cos x \\
\nonumber & - \frac{\sin x}{2 x} \ln  \frac{\sin x}{x}  + \frac{1}{4 x}\sin x -\frac{1}{2} \cos x \ln x \\
& - \frac{\cos x}{4 x} \text{Si}(2 x) -\frac{\sin x}{4 x}
\text{Cin}(x) \bigg] + O[(n-1)^2], \label{eqn:Bender_theta}
\end{align}
where $\text{Si}(x)$ and $\text{Cin}(x)$ are integrals of sine and
cosine defined respectively by \citep[see,
e.g.,][]{Math_handbook}:
\begin{equation}
\text{Si}(x) = \int_{0}^{x} \frac{\sin t}{t} dt,
\end{equation}
\begin{equation}
\text{Cin}(x) = \int_{0}^{x} \frac{1 - \cos t}{t} dt,
\end{equation}
and the normalised radius $\hat{\xi}_{\text{B}}(n)$ as a function of $n$
up to the second order in $n-1$:
\begin{align}
\nonumber \hat{\xi}_{\text{B}}(n) & = \pi + \left(\frac{1}{2} \pi  \ln 2 - \frac{3 \pi}{4}+\frac{1}{2} \pi  \ln \pi + \frac{\text{Si}(2 \pi)}{4}\right) (n-1) \\
\nonumber & \quad + 0.24222 (n-1)^2 + O[(n-1)^3] \\
\label{eqn:Bender_xi} & \approx \pi + 0.885273956 (n-1) + 0.24222
(n-1)^2 + O[(n-1)^3].
\end{align}

Similar to the case of $\hat{\theta}_{\text{S}}(x)$, owing to the
presence of the terms involving $ \ln \sin x$,
$\hat{\theta}_{\text{B}}(x)$ is complex-valued for $x
> \pi$, while the physical interval of interest extends beyond
$\pi$ for $n>1$. In such a case, we still take the real part of
the approximant by means of analytic continuation in the present
paper.

\section{Scaled Delta Expansion Method} \label{sec:homology_perturbation}
In Sections \ref{sec:homology_invariance} and
\ref{sec:literature_review1}, we have introduced the scale
invariance symmetry of the LEE and reviewed previous attempts to
solve the LEE perturbatively by the DEM. In this section, we
incorporate the idea of scale invariance into the DEM, in order to
arrive at uniform approximants of the polytrope function and the
radius with an $n$-dependent scale transformation. We derive in
detail the perturbation schemes about $n=0$ and $n=1$,
respectively. By the end of the section, we compare the numerical
results obtained from SDEM and DEM, and see that SDEM is able to
yield more accurate results through resummation of the
perturbation series obtained previously by
\citet{Bender_delta_expansion} and \citet{Seidov_LEE}.

\subsection{Scaled LEE}
As clearly shown in the LEE \eqref{eqn:LEE}, the equation becomes
singular at the zeros of the polytrope function owing to the
presence of the term $[\theta(x)]^n$ except for cases with $n$
being an integer. Therefore, in order to establish an accurate
perturbation scheme valid in the entire physical domain where
$\theta(x) \ge 0$, it is essential to capture the
location of the $n$-dependent singular point associated with the
radius of the polytrope function. As could be seen from the
exact solutions at $n=0$ and $n=1$, the normalised radius $\hat{\xi}$ of
moves from $\sqrt{6}$ to $\pi$
as $n$ increases from $0$ to $1$. As a matter of fact, $\hat{\xi}$
as well as the the physical domain of the LEE increase
monotonically as a function of $n$ for $0 \le n < 5$. However,
previous perturbative analyses by \citet{Bender_delta_expansion}
and \citet{Seidov_LEE} have omitted the $n$-dependency of
$\hat{\xi}$ during the evaluation of the polytrope function
$\hat{\theta}$. They expanded the normalised polytrope function
about the exact known result at $n=0$ (or $n=1$), but such an
expansion becomes invalid when $x$ crosses the singular point of
the unperturbed polytrope where $x
> \sqrt{6}$ (or $x > \pi$). As a result, in both equations
(\ref{eqn:Seidov_theta}) and  (\ref{eqn:Bender_theta}), the
approximants of $\hat{\theta}(x)$ obtained by DEM, are
ill-defined in real for $x> \sqrt{6}$ and $x> \pi$, respectively.
Here, we properly take this issue into account via an
$n$-dependent scale transformation.

First of all, in order to keep the physical interval of definition
unchanged throughout the course of perturbation, we define an
alternative length scale $z$ in the light of the scale invariance
symmetry  shown  in equation \eqref{eqn:polytrope_radius}:
%described in Theorem \ref{theorem}:
\begin{equation}
\label{eqn:def_scale} x = S(n)^{(n-1)/2} z,
\end{equation}
where the scale factor $S(n)$ is determined by requiring the
the radius to be $z = \pi$ in the new scale we introduce. As
a result, a new differential equation is generated:
\begin{equation}
\mathcal{L}_{z} \left[ S(n) \Theta(z) \right] + \left[S(n)
\Theta(z) \right]^{n} = 0, \label{eqn:LEE_z}
\end{equation}
where $\Theta(z) \equiv \theta(S(n)^{(n-1)/2} z)$ is coined here
as the scaled polytrope function. Equation \eqref{eqn:LEE_z},
hereafter referred to as the scaled LEE (SLEE), is identical to
the conventional LEE \eqref{eqn:LEE} except for an extra scale
factor $S(n)$. Therefore, we obtain another polytrope function
$\theta(z) = S \Theta(z)$ satisfying the LEE under an unconventional
initial condition $ \theta(0) = S$.
Such a scale transformation property of LEE, which is equivalent
to adoption of another density scale $\rho_{0}$ by equation
(\ref{eqn:polytrope_radius}), is commonly referred to as homology
invariance in mathematical texts \citep{Horedt_homology,
Sharaf_homology}, and is summarised in the following theorem:
%\citep{Horedt_homology, Sharaf_homology}:
\begin{theorem}
\label{theorem} Let $n$ be the polytropic index, and $\theta(x)$
a solution to the LEE of polytropic index $n$. For any positive real
number $S$, $S \theta( S^{(n-1)/2} x)$ is also a solution to the
LEE of the same polytropic index $n$.
\end{theorem}

Our original target is to solve for the normalised polytrope
function $\hat{\theta}(x)$ and the associated radius
$\hat{\xi}$ of the LEE \eqref{eqn:LEE}. Accordingly, we look for
the solution of the scaled LEE \eqref{eqn:LEE_z} with the
following requirements, namely, (1) $\Theta(z=0) = 1$; (2)
$\Theta'(z=0)=0$ and (3) $\Theta(z=\pi)=0$. As the SLEE
\eqref{eqn:LEE_z} is a second order ordinary differential
equation, these three requirements in general cannot be satisfied
simultaneously. Therefore, we have to look for a suitable value of
the scale factor $S$ for each value of $n$. In other words, we
have to solve the following  eigenvalue problem:
\begin{equation}
\label{eqn:homology_perturbation_problem} \frac{1}{z^2}
\frac{d}{dz} \bigg( z^2 \frac{d \Theta}{d z} \bigg) =- S(n)^{n-1}
\Theta^{n},
\end{equation}
where $\Theta(z)$ and $S(n)$ are  considered as the eigenfunction
and the eigenvalue respectively, subject to the three requirements
mentioned above. Once $\Theta(z)$ and $S(n)$ are found, we can
invoke  Theorem \ref{theorem} to yield the solutions of
$\hat{\theta}(x)$ and $\hat{\xi}(n)$:
\begin{align}
 &  \hat{\theta}(x) = \Theta(S^{(1-n)/2} x)=\Theta(\pi x/\hat{\xi}(n)), \label{eqn:tt}\\
 & \hat{\xi}(n) =  \pi S(n)^{(n-1)/2}. \label{eqn:xi_S_pi}
\end{align}

Before delving into the details of the solution of the SLEE
\eqref{eqn:homology_perturbation_problem}, we would like to
digress slightly and draw the attention of the readers  to the
similarity among the method of harmonic balance (HB) \citep[see,
e.g.,][]{nayfeh1995nonlinear,marinca2012nonlinear}, the method of
multiple scale analysis (MSA) \citep[see,
e.g.,][]{bender1999advanced} and the SDEM developed here. In the
methods of HB and MSA, which were devised to study non-linear
oscillations, it is customary to define an alternative time scale,
so that the period of oscillation is fixed, say, at $2\pi$ in the
new time scale regardless of the strength of the nonlinear
coupling, and the time scale transformation is determined by
demanding the absence of secular terms (i.e., terms resonantly
coupled with the unperturbed state).
%\citep{nayfeh1995nonlinear,bender1999advanced,
%marinca2012nonlinear}.
In the present case, we define an auxiliary length scale, in a way
that the radius is fixed at $z=\pi$ regardless of
the polytropic index $n$, which in turn determines the scale factor
$S$.  In both methods of HB and MSA, the introduction of the
auxiliary time scale is able to capture the frequency shift due to
the non-linear coupling, while eliminating the possible secular terms.
In our SDEM for the LEE, we shall see that the length scale
transformation reveals the $n$-dependence of the radius, at which
a branch point singularity of the polytrope function occurs,
and thereby getting rid of complex-valued solutions.

\subsection{SDEM about $n=0$}
We have successfully transformed the solution of the LEE
\eqref{eqn:LEE} from an initial value problem, where the initial
conditions shown in equation \eqref{eqn:IC9}
%$\hat{\theta}(0)=1$ and $\hat{\theta}'(0)=0$
are imposed, into an eigenvalue problem
subject to the three boundary conditions aforementioned. These
three conditions in principle allow us to determine the scaled
polytrope function $\Theta(z)$ and the scale factor $S(n)$
uniquely. However, there is no closed form solution to this
eigenvalue problem except for $n=0$ and $1$. In the following, we employ
the DEM proposed by \citet{Bender_delta_expansion} to expand the
non-linear term in the SLEE
(\ref{eqn:homology_perturbation_problem}) about the two known
analytical solutions with $n=0$ and $1$. In
contrast to the DEM calculations outlined in Section
\ref{sec:literature_review1}, we have to consider the $n$-dependency of the scale
factor $S(n)$ and the additional boundary condition
$\Theta(z=\pi)=0$.

First of all, we consider the expansion about the incompressible
limit at $n=0$. We expand $\Theta(z)$ and the scale factor
$S(n)$ in the SLEE (i.e., equation \eqref{eqn:LEE_z} or
(\ref{eqn:homology_perturbation_problem})) in series in $n$,
\begin{align}
&\Theta(z) = \Theta_{0}^{(0)}(z) + n \Theta_{0}^{(1)}(z)+ n^2
\Theta_{0}^{(2)}(z) + n^3 \Theta_{0}^{(3)}(z) + O[n^4],\\
&S(n) = S_{0}^{(0)} + n S_{0}^{(1)}  + n^2 S_{0}^{(2)}  + n^3
S_{0}^{(3)}  + O[n^4],
\end{align}
where the subscript $0$ and the superscript $i$ in parenthesis of
$\Theta_{0}^{(i)}(z)$ and $S_{0}^{(i)}$ ($i=0,1,2,\ldots$)
respectively denote the unperturbed value of $n$ and the order of
perturbation, and then solve the resultant equations order by
order in $n$. The equations resulting from the leading three
orders read:
\begin{align}
& S_{0}^{(0)} \mathcal{L}_{z} \Theta_{0}^{(0)}(z) = - 1,\\
& S_{0}^{(0)} \mathcal{L}_{z} \Theta_{0}^{(1)}(z) = - S_{0}^{(1)} \mathcal{L}_{z}
\Theta_{0}^{(0)}(z) - \ln \Theta_{0}^{(0)}(z) -\ln S_{0}^{(0)}, \\
\label{eqn:2nd_order_differential_eqn}
& \nonumber S_{0}^{(0)} \mathcal{L}_{z} \Theta_{0}^{(2)}(z) \\
\nonumber = & -\frac{1}{2} \ln ^2 S_{0}^{(0)} - \ln  S_{0}^{(0)} \ln \Theta_{0}^{(0)}(z) - \frac{1}{2} \ln ^2 \Theta_{0}^{(0)}(z) \\
& -\frac{S_{0}^{(1)}}{S_{0}^{(0)}}
-\frac{\Theta_{0}^{(1)}(z)}{\Theta_{0}^{(0)}(z)} - S_{0}^{(1)}
\mathcal{L}_{z} \Theta_{0}^{(1)}(z) - S_{0}^{(2)} \mathcal{L}_{z}
\Theta_{0}^{(0)}(z),
\end{align}
which are subject to the following three boundary conditions:
\begin{equation}\label{bc3}
\Theta_{0}^{(i)}(z=0) = \delta_{0i}; \quad \frac{d
\Theta_{0}^{(i)}}{d z}|_{z=0} = 0; \quad {\rm and } \quad
\Theta_{0}^{(i)}(z=\pi) = 0
\end{equation}
for $i=0,1,2,\ldots$.

In general, we find that in each order of the perturbation
equations listed above  the following  inhomogeneous equation
arises:
\begin{equation}
\frac{1}{z^2} \frac{d}{dz}\bigg( z^2 \frac{d y}{d z} \bigg) = f(z),
\end{equation}
where $y(z)$ is the function to be determined, and $f(z)$ is a
given inhomogeneous term. Using method of variation of parameters
and the two independent solutions to the associated homogeneous
equation, namely $1$ and $1/z$, we obtain the general solution to
$y(z)$:
\begin{equation}
\label{eqn:variation_of_parameters_n0}
y(z) = \int^{z}_{0} (t - \frac{t^2}{z})  f(t) dt+C,
\end{equation}
where $C$ is an integration constant. In particular, if $C=0$,
then the initial conditions $y(0)=0$ and $y'(0)=0$ are satisfied,
as long as the inhomogeneous term $f(z)$ is analytic at $z=0$.

From the perturbation equations, the boundary conditions in
equation \eqref{bc3} and the general solution
\eqref{eqn:variation_of_parameters_n0} to these perturbation
equations, we obtain  the solutions for $S_{0}^{(i)}$ and
$\Theta_{0}^{(i)} (z)$ for $i=0,1,2$:
\begin{equation}
\label{eqn:S00_solution}
S_{0}^{(0)} = \frac{\pi^2}{6},
\end{equation}
\begin{equation}
\label{eqn:theta00_solution} \Theta_{0}^{(0)} (z) = 1 -
\frac{z^2}{\pi^2},
\end{equation}
\begin{equation}
\label{eqn:S01_solution}
S_{0}^{(1)} = \frac{7 \pi ^2}{18}-\frac{2 \pi ^2}{3} \ln 2+\frac{\pi ^2}{6} \ln \frac{\pi ^2}{6} \approx 0.0961380293559132,
\end{equation}
\begin{align}
\nonumber \Theta_{0}^{(1)}(z) = & -4 + \frac{4 z^2}{\pi^2} \left(1 - \ln 2 \right)
+\left(3-\frac{2 \pi }{z}-\frac{z^2}{\pi ^2}\right) \ln \left(1-\frac{z}{\pi}\right)\\
& +\left(3+\frac{2 \pi }{z}-\frac{z^2}{\pi ^2}\right) \ln
\left(1+\frac{z}{\pi }\right),\label{eqn:theta01_solution}
\end{align}
\begin{align}
\nonumber S_{0}^{(2)} & = -\frac{287 \pi ^2}{54}+\frac{7 \pi ^4}{18} + \frac{10}{9} \pi ^2 \ln 2 +
\frac{4}{3} \pi ^2 \ln ^2 2 +\frac{5}{9} \pi ^2 \ln \frac{\pi ^2}{6} \\
\nonumber & \quad  + \frac{1}{12} \pi ^2 \ln ^2 \frac{\pi ^2}{6} -\frac{2}{3} \pi ^2 \ln 2 \ln \frac{\pi ^2}{6} \\
& \approx 0.01271356406241326,\label{eqn:S02_solution}
\end{align}
\begin{align}
\Theta_{0}^{(2)}(z) = & 40+\frac{7 \pi ^2}{3}+8 \ln 2-14 \ln ^2 2 \nonumber \\
& + \left(\frac{7 \pi^2}{3}- 40 +32 \ln 2 -8 \ln ^2 2 \right) \frac{z^2}{\pi^2} \nonumber \\
& + \left[-23+\frac{20 \pi }{z}+\frac{3 z^2}{\pi ^2}+ \big(14 -\frac{10 \pi}{z}-\frac{4 z^2}{\pi ^2}\big)\ln 2 \right]\nonumber \\
& \quad \times \ln \left(1-\frac{z}{\pi}\right) + \left(\frac{3}{2}-\frac{\pi }{z}-\frac{z^2}{2 \pi ^2}\right) \ln ^2\left(1-\frac{z}{\pi }\right) \nonumber \\
& + \left[-23-\frac{20 \pi }{z}+\frac{3 z^2}{\pi ^2}+\big( 14+\frac{10 \pi}{z}-\frac{4 z^2}{\pi ^2}\big) \ln 2 \right] \nonumber \\
& \quad \times \ln \left(1+\frac{z}{\pi}\right) + \left(\frac{3}{2}+\frac{\pi }{z}-\frac{z^2}{2 \pi^2}\right) \ln ^2\left(1+\frac{z}{\pi }\right) \nonumber \\
& + \left(1-\frac{z^2}{\pi ^2}\right) \ln \left(1-\frac{z}{\pi }\right) \ln \left(1+\frac{z}{\pi }\right)\nonumber \\
& + \left(\frac{14 \pi }{z}-14\right) \text{Li}_2\left(\frac{\pi -z}{2 \pi }\right)\nonumber \\
& + \left(-\frac{14 \pi }{z}-14\right) \text{Li}_2\left(\frac{\pi
+z}{2 \pi}\right).\label{eqn:theta02_solution}
\end{align}
Here $\text{Li}_2(z)$, called the dilogarithm (or polylogarithm of
order $2$), is defined by \citep[see,
e.g.,][]{Math_handbook,Lewin_polylog}:
\begin{equation}
\text{Li}_2 (z) = \int_{0}^{z} \frac{ - \ln(1-t) }{t} dt,
\end{equation}
and for $|z|<1$ it is also given by:
\begin{equation}
\text{Li}_2 (z)  = \sum^{\infty}_{k=1} \frac{z^k}{k^2}.
\end{equation}
In particular, the following formulae for the dilogarithm function
are useful in the evaluation of  $S_{0}^{(2)}$ \citep[see,
e.g.,][]{Math_handbook,Lewin_polylog}:
\begin{align}
& \text{Li}_2(\frac{1}{2}) = \frac{\pi ^2}{12}-\frac{\ln ^2 2}{2}, \\
& \text{Li}_2(1) = \zeta(2) = \frac{\pi^2}{6},
\end{align}
where $\zeta(s)$ is the Riemann zeta function defined by the
infinite series:
\begin{equation}
\zeta(s) = \sum_{k=0}^{\infty} \frac{1}{k^{s}}.
\end{equation}

Two remarks are in order. First, from $\Theta_{0}^{(1)}(z)
\leq 0$ for all $z \in [0,\pi]$, we see that the density decreases
with increasing polytropic index $n$, when scaled to a common
radius. Also, from equation (\ref{eqn:xi_S_pi}) and that
$S_{0}^{(0)} \ln S_{0}^{(0)} - S_{0}^{(1)} \approx 0.7225 > 0$, we
conclude that the normalised radius $\hat{\xi}$ increases with the
polytropic index $n$.  It agrees with our experience that the
larger the polytropic index $n$, the more extended the density
distribution, and the larger is the physical radius.

We have also evaluated the third order correction $S_{0}^{(3)}$ in
a similar fashion:
\begin{align}
\nonumber S_{0}^{(3)} = & \frac{18113}{324}\pi^2 -\frac{46}{27}\pi^4-\bigg(\frac{212}{27} \pi^2 + \frac{14}{9} \pi^4 \bigg) \ln 2\\
\nonumber &  -\frac{68}{9} \pi ^2 \ln ^2 2 -\frac{16}{9} \pi ^2 \ln ^3 2 -\frac{47}{3} \pi ^2 \zeta (3) +  \bigg( - \frac{257}{54} \pi ^2 \\
\nonumber & +\frac{7}{18} \pi ^4 +\frac{4}{9} \pi ^2 \ln 2 + \frac{4}{3} \pi ^2 \ln ^2 2\bigg) \ln \frac{\pi ^2}{6} \\
\nonumber & + \bigg( \frac{13}{36} \pi^2 - \frac{1}{3} \pi^2 \ln 2 \bigg)  \ln ^2 \frac{\pi ^2}{6} +\frac{1}{36} \pi ^2 \ln^3 \frac{\pi ^2}{6}  \\
\label{eqn:S03_solution} & \approx 0.002686585492882264.
\end{align}
We do not record the third order correction of the polytrope
function $\Theta_{0}^{(3)}(z)$ here, for its enormous algebraic
complexity and limited usefulness.

\subsection{SDEM about $n=1$}
The perturbation scheme for the case $n=1$ is similar to that in
the case $n=0$. The starting point is still the simultaneous
expansion of both $\Theta(z)$ and $S(n)$ in power series of $n-1$:
% We expand the non-linear terms in
%Eq. (\ref{eqn:homology_perturbation_problem}) in a power series of
%$n-1$, and then solve the resultant equations order by order, in
%the same spirit as \citet{Bender_delta_expansion}.  We write:
\begin{align}
&\Theta(z) = \Theta_{1}^{(0)}(z) + (n-1) \Theta_{1}^{(1)}(z)+
(n-1)^2 \Theta_{1}^{(2)}(z) + O[(n-1)^3],\\
%\end{equation}
%\begin{equation}
&S(n) = S_{1}^{(0)} + (n-1) S_{1}^{(1)}  + (n-1)^2 S_{1}^{(2)}  +
O[(n-1)^3],
\end{align}
where the subscript 1 of $\Theta_{1}^{(i)}(z)$ and $S_{1}^{(i)}$
($i=0,1,2,\ldots$) indicates the unperturbed value of $n$. When
these expansions are inserted into the SLEE \eqref{eqn:LEE_z},
coupled perturbation equations can be obtained order by order,
with the leading three differential equations given explicitly as
follows:
\begin{align}
& {\cal L}_z \Theta_{1}^{(0)}+ \Theta_{1}^{(0)} =0, \label{eqn:theta10_definition} \\
\label{eqn:theta11_definition}
& {\cal L}_z \Theta_{1}^{(1)}+ \Theta_{1}^{(1)} =
-\Theta_{1}^{(0)} \ln S_{1}^{(0)} - \Theta_{1}^{(0)} \ln
\Theta_{1}^{(0)} , \\
& {\cal L}_z \Theta_{1}^{(2)}+\Theta_{1}^{(2)} \nonumber \\
= & -\frac{1}{2} \Theta_{1}^{(0)} \ln ^2 S_{1}^{(0)} - \Theta_{1}^{(0)} \ln S_{1}^{(0)} \ln \Theta_{1}^{(0)} -\frac{1}{2} \Theta_{1}^{(0)} \ln ^2 \Theta_{1}^{(0)} \nonumber \\
& -\frac{S_{1}^{(1)} \Theta_{1}^{(0)}}{S_{1}^{(0)}} -
\Theta_1^{(1)} \ln S_{1}^{(0)} -\Theta_1^{(1)}  -
\Theta_1^{(1)}  \ln
\Theta_{1}^{(0)}, \label{eqn:theta12_definition}
\end{align}
under the boundary conditions:
\begin{equation}\label{eqn:boundary_conditions_n1}
\Theta_{1}^{(i)}(z=0) = \delta_{0i}; \quad \frac{d
\Theta_{1}^{(i)}}{d z}|_{z=0} = 0; \quad {\rm and } \quad
\Theta_{1}^{(i)}(z=\pi) = 0
\end{equation}
for $i=0,1,2,\ldots$.

Once again we see that these perturbation equations are reducible
to a common form:
\begin{equation}
\frac{d^2 y}{d z^2} + \frac{2}{z} \frac{d y}{dz} + y = f(z),
\end{equation}
which can be identified as the zeroth order spherical Bessel
equation in the absence of the inhomogeneous term $f(z)$. It is
well known that the two independent solutions to the associated
homogeneous equation are $\sin z / z$ and $\cos z / z$. By
variation of parameters, the general solution to $y(z)$ is given
by:
\begin{equation}
\label{eqn:variation_of_parameters_n1} y(z) = \frac{\sin z}{z}
\int^{z}_{0} t f(t) \cos t dt - \frac{\cos z}{z} \int^{z}_{0} t
f(t) \sin t dt+C,
\end{equation}
where $C$ is an integration constant. In particular,  the initial
conditions $y(0)=0$ and $y'(0)=0$ are readily satisfied if $C=0$
on condition that the inhomogeneous term $f(z)$ is analytic at
$z=0$.

Although the perturbation equations for the cases $n=0$ and $n=1$
are obtained from the same principle, there is a crucial
difference between these two sets of equations. In the former
case, the expansion coefficients $S_{0}^{(0)}, \ldots,S_{0}^{(i)}$
explicitly appear in the governing equation of $\Theta_{0}^{(i)}$
for $i=0,1,2,\dots$. In contrast, in the latter case, the
governing equation of $\Theta_{1}^{(i)}$ depends on the expansion
coefficients $S_{1}^{(0)}, \ldots,S_{1}^{(i-1)}$ explicitly and
does not involve $S_{1}^{(i)}$. As a result, the zeroth order
perturbation differential equation,
\eqref{eqn:theta10_definition}, is independent of $S_{1}^{(0)}$
and is indeed the zeroth order spherical Bessel equation with the
following solution:
\begin{equation}
\label{eqn:theta10_solution} \Theta_{1}^{(0)}(z) = \frac{\sin
z}{z},
\end{equation}
where the boundary conditions in equation
\eqref{eqn:boundary_conditions_n1} have already been imposed.

We note that in equation (\ref{eqn:variation_of_parameters_n1}),
at the boundary $z = \pi$, $\sin z/z$ vanishes and only the $\cos
z/z$ term contributes. Hence, upon identifying the inhomogeneous
term on the right hand side in equation
(\ref{eqn:theta11_definition}), the value of $S_{1}^{(0)}$ can be
found by equating  the second integral in equation
(\ref{eqn:variation_of_parameters_n1}) to zero:
\begin{equation}
\label{eqn:S10_solution} S_{1}^{(0)} = 2 \pi  \exp\bigg[
\frac{\text{Si}(2 \pi )}{2 \pi }-\frac{3}{2} \bigg].
\end{equation}
Using equations \eqref{eqn:theta10_solution} and
\eqref{eqn:S10_solution} as the input to the first-order
perturbation equation \eqref{eqn:theta11_definition},
we determine $\Theta_{1}^{(1)}(z)$:
\begin{align}
\Theta_{1}^{(1)}(z) & = \frac{\sin z}{z} \bigg[ 1-\frac{\ln(2\pi)}{2}-\frac{\text{Si}(2 \pi )}
{4 \pi } - \frac{\text{Cin}(2 z)}{4} \nonumber \\
& +\frac{\ln z}{2}-\frac{1}{2} \ln(\sin z) \bigg] + \frac{\cos z}{z} \bigg[ \frac{1}{2} z \ln (2 \pi) \nonumber \\
& -\frac{1}{2} z \ln z+\frac{z \text{Si}(2 \pi )}{4 \pi
}-\frac{\text{Si}(2 z)}{4} + \frac{1}{2} \int_{0}^{z} \ln \sin t
dt \bigg].\label{eqn:theta11_solution}
\end{align}
%\end{equation}

Once again we observe that $\Theta_{1}^{(1)}(z) \leq 0$ for all $z
\in [0,\pi]$, which implies that the density of the polytrope at a
fixed scaled variable $z$ decreases monotonically with increasing
$n$ at $n=1$. Moreover, from equation ($\ref{eqn:xi_S_pi}$) and
that $\ln S_{1}^{(0)} \approx 0.5636 > 0$ , we can see that the
normalised radius $\hat{\xi}$ increases with increasing $n$ as well. It
again confirms our intuition that polytropic density tail
lengthens with increasing polytropic index $n$.

On the other hand, analytical solution to $\Theta_{1}^{(2)}(z)$
and $S_{1}^{(1)}$ would involve cumbersome integrals of the
special functions $\text{Si}$ and $\text{Cin}$.
%and it is of limited practical
%usefulness to express the solution as integrals of these special
%functions.
Instead, we determine $S_{1}^{(1)}$ by numerically
integrating the second integral in equation
(\ref{eqn:variation_of_parameters_n1}) from $0$ to $\pi$ and then
equating it to zero. The solution is given by:
\begin{equation}
\label{eqn:S11_solution}
S_{1}^{(1)} \approx 0.13168015025423085.
\end{equation}

\subsection{Numerical results}
Using $\Theta(z)$ and $S(n)$ obtained perturbatively as outlined
above, we can find the normalised polytrope function
$\hat{\theta}(x)$ and the normalised radius of the standard LEE,
$\hat{\xi}$, from equations \eqref{eqn:tt} and
\eqref{eqn:xi_S_pi}. Hence, the physical radius $R$ of a polytrope
can be obtained from equation \eqref{eqn:polytrope_radius} by
noting that $\theta(0)^{(n-1)/2} \xi = \hat{\xi}$ and is given by:
\begin{equation} \label{eqn:polytrope_radius_R}
R  = \sqrt{\frac{K (n+1)}{4 \pi G}} \rho_{c}^{(1-n)/(2n)} \hat{\xi}. \\
\end{equation}
In the following we show the accuracy of the
numerical results of the SDEM and compare them with those obtained
from the DEM by \citet{Seidov_LEE} and
\citet{Bender_delta_expansion}.

In general, we can truncate the SDEM perturbation series for the
normalised radius $\hat{\xi}_{p}$ (see equation \eqref{eqn:xi_S_pi}),
where $p=0$ or $1$ respectively signifies expansion about $p=0$ or
$1$, at the $j$-th order ($j=0,1,2,\ldots$) in $n-p$ and denote
such a partial sum as:
\begin{equation}
\label{eqn:xi_local} [\hat{\xi}_{p}]_j (n) = \pi [S_{0}^{(0)} + (n-p)
S_{0}^{(1)} +\ldots+ (n-p)^j S_{0}^{(j)} ]^{(n-1)/2}.
\end{equation}
In the present paper explicit expressions for $S_{0}^{(0)}$,
$S_{0}^{(1)}$, $S_{0}^{(2)}$, $S_{0}^{(3)}$, $S_{1}^{(0)}$, and $S_{1}^{(1)}$
have been derived in equations (\ref{eqn:S00_solution}),
(\ref{eqn:S01_solution}), (\ref{eqn:S02_solution}), (\ref{eqn:S03_solution}),
(\ref{eqn:S10_solution}) and (\ref{eqn:S11_solution}),
respectively.

Furthermore,  we  insert $[\hat{\xi}_{p}(n)]_j$ into equation
\eqref{eqn:tt} to evaluate the $k$-th order ($k=0,1,2,\ldots$)
partial sum  of the polytrope function $\hat{\theta}_{p}(x)$:
\begin{equation}
\label{eqn:theta_local} [\hat{\theta}_{p}]_j^k (x) =
\Theta_{p}^{(0)}(z_j) + (n-p) \Theta_{p}^{(1)} (z_j)+ \ldots +
(n-p)^k \Theta_{p}^{(k)} (z_j),
\end{equation}
with $z_j=\pi x / [\hat{\xi}_{p}]_j(n)$. In this paper we have
explicitly determined  $\Theta_{0}^{(0)}(z)$,
$\Theta_{0}^{(1)}(z)$, $\Theta_{0}^{(2)}(z)$,
$\Theta_{1}^{(0)}(z)$ and $\Theta_{1}^{(1)}(z)$ as given by
equations (\ref{eqn:theta00_solution}),
(\ref{eqn:theta01_solution}), (\ref{eqn:theta02_solution}),
(\ref{eqn:theta10_solution}) and (\ref{eqn:theta11_solution})
respectively.

We gauge the accuracy and the convergence of the partial sums
given in equations \eqref{eqn:xi_local} and
\eqref{eqn:theta_local} against the numerically exact solution. As
an overview, it could be observed that the accuracy of the SDEM
approximants improves uniformly order by order throughout the
entire physical interval extending from the centre to the surface of a
polytrope. Also, the SDEM approximants are more accurate than the
DEM counterpart, which implies the introduction of the scale
transformation entails an effective series resummation.

\begin{figure}
    \includegraphics[width = 0.49 \textwidth]{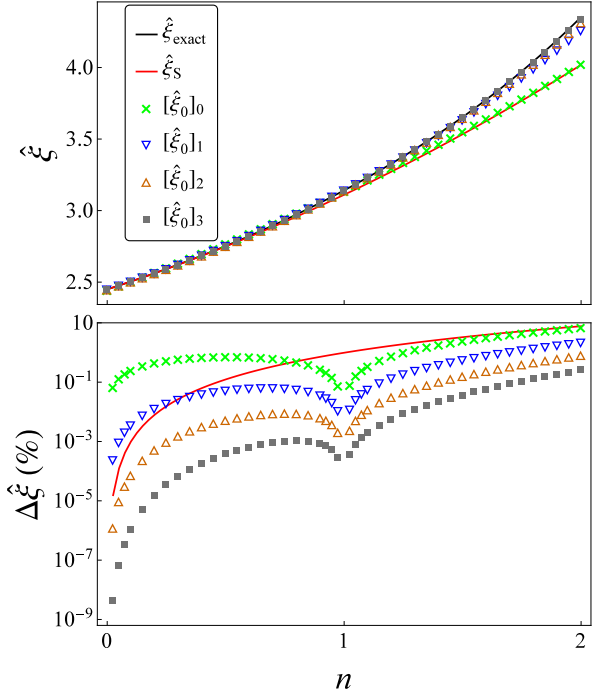}
    \caption{\label{fig:xi_Seidov_SDEM}
    In the upper panel, the value of the normalised radius $\hat{\xi}$ is plotted as a function of the
    polytropic index $n$
    for the numerically exact solution $\hat{\xi}_{\text{exact}}$ (black line),
    the SDEM approximants about $n=0$, $[\hat{\xi}_{0}]_{j}$ in equation (\ref{eqn:xi_local}) for $j = 0$
    (green cross), $j=1$ (blue triangle), $j=2$ (orange triangle) and $j=3$ (grey square), and
    the second order DEM approximation about $n=0$, $\hat{\xi}_{\text{S}}$
    (see equation (\ref{eqn:Seidov_xi}), red line).
    The associated error plot is shown in the lower panel, where
    $\Delta\hat{\xi} = |\hat{\xi} - \hat{\xi}_{\text{exact}}|/\hat{\xi}_{\text{exact}} \times 100\%$
    represents the percentage error of the approximant.
    }
\end{figure}

\begin{figure}
    \includegraphics[width = 0.49 \textwidth]{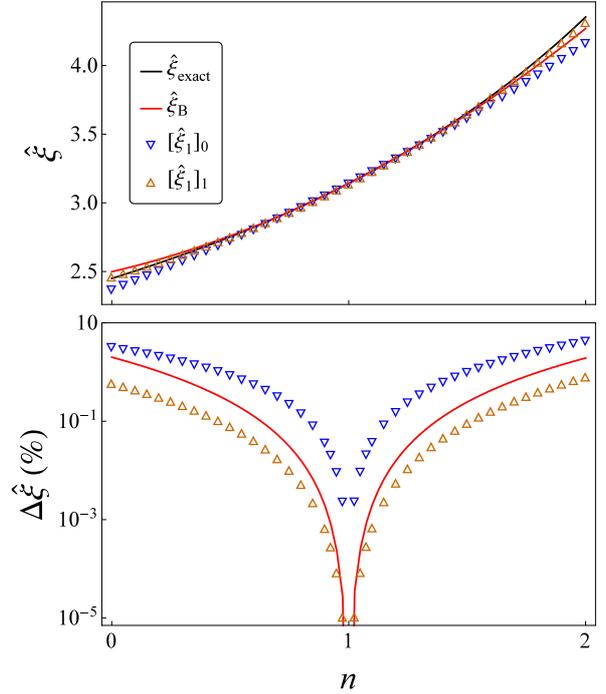}
    \caption{\label{fig:xi_Bender_SDEM}
    In the upper panel, the value of the normalised radius $\hat{\xi}$
    is plotted as a function of the polytropic index $n$
    for the numerically exact solution $\hat{\xi}_{\text{exact}}$ (black line),
    the SDEM approximants about $n=1$, $[\hat{\xi}_{1}]_{j}$ in equation (\ref{eqn:xi_local})
    for $j = 0$ (blue triangle) and $j=1$ (orange triangle), and the second order DEM approximation about
    $n=1$,
    $\hat{\xi}_{\text{B}}$ (see equation (\ref{eqn:Bender_xi}), red line).
    The associated error plot is shown in the lower panel,
    where $\Delta\hat{\xi} = |\hat{\xi} - \hat{\xi}_{\text{exact}}|/\hat{\xi}_{\text{exact}} \times 100\%$
    represents the percentage error of the approximant.}
\end{figure}

In the upper panel of Figure \ref{fig:xi_Seidov_SDEM}, we have
plotted the normalised radius
$\hat{\xi}(x)$ against the polytropic index $n$. In particular,
we compare the numerically exact value of the normalised radius,
$\hat{\xi}_{\text{exact}}$, with the leading four
 perturbative approximations $[\hat{\xi}_{0}]_j$ ($j=0,1,2,3$) obtained from the SDEM about the
incompressible limit $n=0$. As could be seen from the lower panel
of Figure \ref{fig:xi_Seidov_SDEM}, where the percentage error
$\Delta\hat{\xi} \equiv |\hat{\xi} -
\hat{\xi}_{\text{exact}}|/\hat{\xi}_{\text{exact}} \times 100\%$
is shown as a function of $n$ for different approximants, the
accuracy of the SDEM approximants $[\hat{\xi}_{0}]_j$ improves by
two orders of magnitude over the interval $n \in [0,2]$ for each
unit increment in the perturbation order $j$. It suggests that the
proposed SDEM is able to offer rapid and uniform convergence.
Moreover, the dips of the SDEM approximants at $n=0$ and $n=1$ in
this error plot correspond to the fact that the SDEM approximants
are exact at these two points.  It is interesting to note that the
SDEM approximant developed about $n=0$ is also exact at $n=1$.
Such an unexpected result can be understood from equation
\eqref{eqn:xi_local}, which guarantees that $[\hat{\xi}_{0}]_{j}$
is equal to $\pi$, which is the exact value at $n=1$, irrespective
of the value of the scale factor $S$. In fact, this is the reason
why we have chosen the normalisation factor in such a way that the
first zero occurs at $z=\pi$ in the new length scale in equation
\eqref{eqn:xi_S_pi}. It is also worth noting that the second order
SDEM approximant $[\hat{\xi}_{0}]_{2}$ is more accurate than its
DEM counterpart $\hat{\xi}_{\text{S}}$ obtained by
\citet{Seidov_LEE} (see equation (\ref{eqn:Seidov_xi}))
 throughout the entire interval for $n$ in
$[0,2]$. The underlying reason for the high accuracy achievable in
the SDEM is the introduction of the scale factor $S(n)$, which
leads to an effective resummation of the series obtained from
the DEM.

Similarly, in Figure \ref{fig:xi_Bender_SDEM}, we contrast
$\hat{\xi}_{\text{exact}}$ with the leading two
perturbative approximations $[\hat{\xi}_{1}]_j$ ($j=0,1$) obtained from the SDEM about
$n=1$.   As the approximants are
exact at $n=1$, there are dips in the error plot (the lower panel
of Figure \ref{fig:xi_Bender_SDEM}) at $n=1$. Also, the first
order SDEM approximant $[\hat{\xi}_{1}]_{1}$ outperforms the
second order DEM counterpart $\hat{\xi}_{\text{B}}$ (see equation
(\ref{eqn:Bender_xi})) obtained by \citet{Bender_delta_expansion}
throughout the entire interval for $n$ in $[0,2]$. This once again
confirms the power of the scale factor $S(n)$.

\begin{figure}
    \includegraphics[width = 0.49 \textwidth]{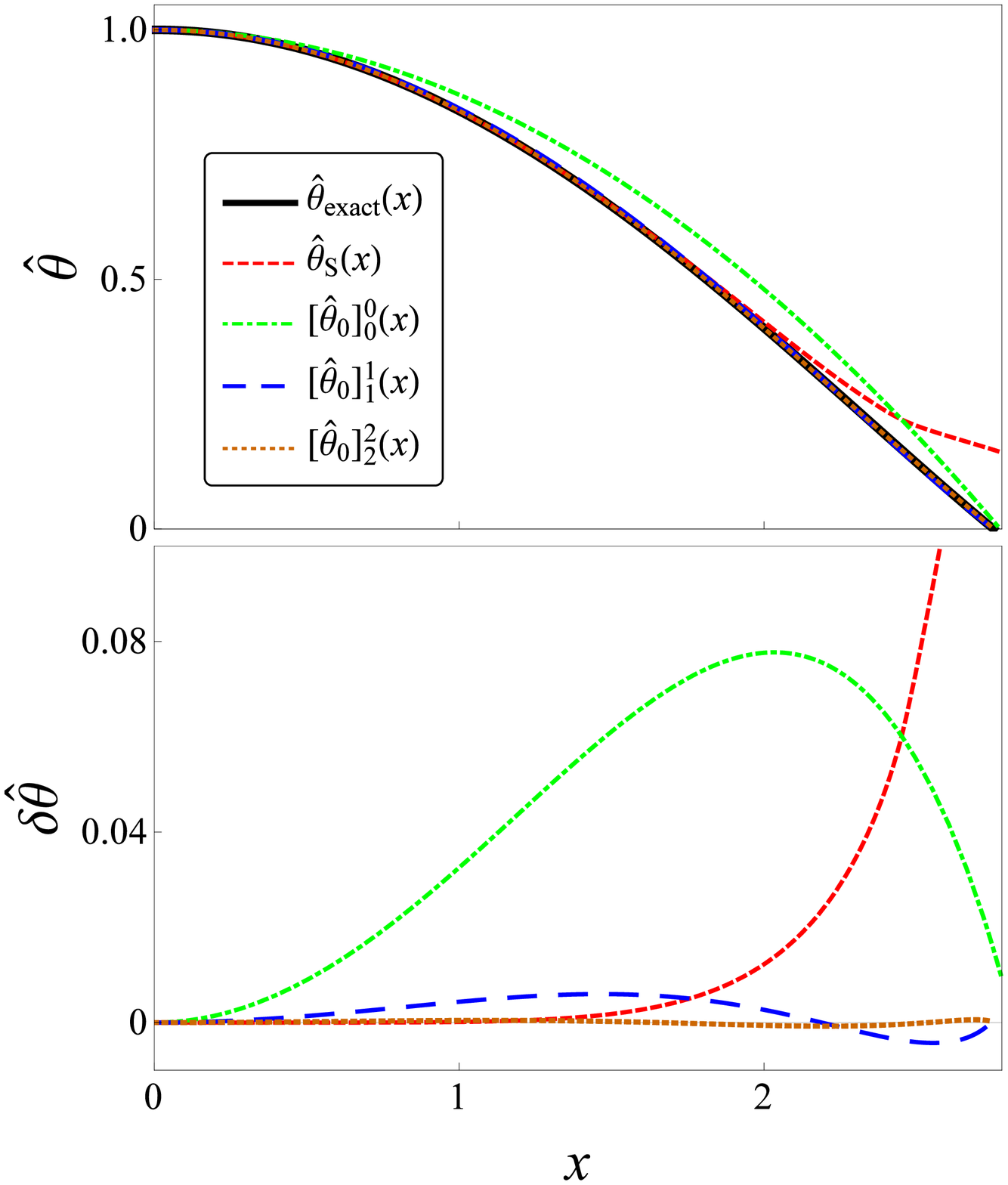}
    \caption{\label{fig:theta_Seidov_SDEM} In the upper panel, for the case of $n=0.5$, we show
    the numerically exact polytrope function $\hat{\theta}_{\text{exact}}(x)$ (black line),
    the  corresponding diagonal SDEM approximants  $ [\hat{\theta}_{0}]_j^j (x)$ about $n=0$
    (see (\ref{eqn:theta_local}))
    for $j=0$ (green dot-dashed), $j=1$ (blue long dashed) and $j=2$ (orange
    dotted), and
    the first order DEM result about $n=0$, $\hat{\theta}_{\text{S}}(x)$
    (see equation (\ref{eqn:Seidov_theta}), red dashed).
    As $\hat{\theta}_{\text{S}}(x)$ is ill-defined in real for $x > \sqrt{6}$,
        we take its real part  in the plot.
    In the lower panel,  the associated deviation from the numerically exact
    solution,
    $\delta \hat{\theta}(x) \equiv \hat{\theta}(x) - \hat{\theta}_{\text{exact}}(x)$ is shown  under the same
    legend.}
\end{figure}

\begin{figure}
    \includegraphics[width = 0.49 \textwidth]{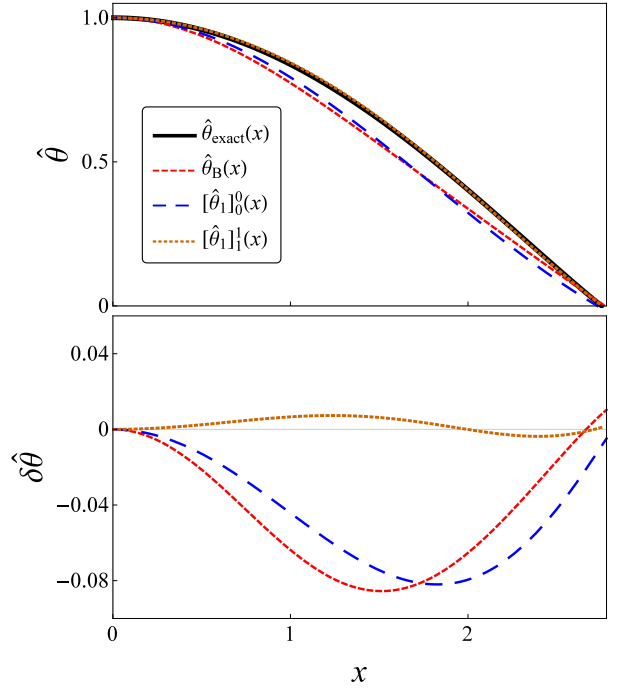}
    \caption{\label{fig:theta_Bender_SDEM}
    In the upper panel, for the case of $n=0.5$, we show
    the numerically exact polytrope function $\hat{\theta}_{\text{exact}}(x)$ (black line),
    the  corresponding diagonal SDEM approximants  $ [\hat{\theta}_{0}]_j^j (x)$ about $n=1$
    (see (\ref{eqn:theta_local}))
     for $j=0$ (blue long dashed) and $j=1$ (orange dotted), and
    the first order DEM result about $n=1$, $\hat{\theta}_{\text{B}}(x)$
    (see equation (\ref{eqn:Bender_theta}), red dashed).
     In the lower panel,  the associated deviation from the numerically exact
    solution,
    $\delta \hat{\theta}(x) \equiv \hat{\theta}(x) - \hat{\theta}_{\text{exact}}(x)$ is shown  under the same
    legend.
        }
\end{figure}

Next, in Figure \ref{fig:theta_Seidov_SDEM}  we study  the
accuracy of the normalised polytrope function $\hat{\theta}(x)$
obtained from the SDEM and the DEM approximants expanded about the
incompressible limit where $n=0$ for a typical case of $n=0.5$. In
the upper panel, the numerically exact polytrope function
$\hat{\theta}_{\text{exact}}(x)$, the leading three diagonal SDEM
approximants  $ [\hat{\theta}_{0}]_j^j (x)$ about $n=0$ (see
equation (\ref{eqn:theta_local})) for $j=0,1,2$, and the first
order DEM result about $n=0$, $\hat{\theta}_{\text{S}}(x)$ (see
equation (\ref{eqn:Seidov_theta})) are shown. Besides, in the
lower panel  the associated deviation from the numerically exact
solution, $\delta \hat{\theta}(x) \equiv \hat{\theta}(x) -
\hat{\theta}_{\text{exact}}(x)$ is also shown. We observe that the
diagonal SDEM approximants $[\hat{\theta}_{0}]_{j}^{j}(x)$ in
general carry bounded error within the entire polytrope. In
particular, the accuracy improves order by order, and the
deviation of the second order diagonal SDEM approximant
$[\hat{\theta}_{0}]_{2}^{2}$ from the exact solution is less than
0.01 in the whole range. On the other hand, the approximant
$\hat{\theta}_{\text{S}}(x)$ obtained from the DEM is also
accurate for small $x$, but the accuracy declines rapidly when $x$
approaches $\sqrt{6}$. In fact, $\hat{\theta}_{\text{S}}(x)$
becomes complex-valued for $x$ greater than $\sqrt{6}$,  the
normalised radius of an incompressible polytrope, and as a
consequence we have to take its real part in the plot.

We perform a similar analysis in Figure
\ref{fig:theta_Bender_SDEM} by contrasting
$\hat{\theta}_{\text{exact}}(x)$ with the diagonal SDEM and the
DEM approximants expanded  about $n=1$ for the case $n=0.5$. We
see that the accuracies of the zeroth order diagonal SDEM
approximant $[\hat{\theta}_{1}]_{0}^{0}$ and the first order DEM
approximant $\hat{\theta}_{\text{B}}(x)$  (see equation
(\ref{eqn:Bender_theta})) are comparable. On the other hand, the
first order diagonal SDEM approximant $[\hat{\theta}_{1}]_{1}^{1}$
carries a small error (less than 0.01) throughout the entire
polytrope and is overall much better than
$\hat{\theta}_{\text{B}}(x)$.

In contrast to the case studied in Figure
\ref{fig:theta_Seidov_SDEM}, $\hat{\theta}_{\text{B}}(x)$ shown in
Figure \ref{fig:theta_Bender_SDEM} is still well behaved near the
surface of the polytrope. Such a qualitative difference can be
understood as follows. In the case considered in Figure
\ref{fig:theta_Bender_SDEM}, the polytropic index is 0.5, which is
less than the perturbation centre $n=1$. As mentioned above,
$\hat{\xi}$ increase with increasing polytropic index. The radius
of the polytrope (with $n=0.5$) considered in Figure
\ref{fig:theta_Bender_SDEM} is thus less than that of the
unperturbed one (with $n=1$).  As a result,
$\hat{\theta}_{\text{B}}(x)$ remains real in the entire polytrope.
If, instead, a polytrope with a polytropic index greater than
unity is considered, $\hat{\theta}_{\text{B}}(x)$ still becomes
complex valued near the surface of the polytrope where $x$ is
larger than $\pi$, the normalised radius of a polytrope with
$n=1$, as in  the case discussed in Figure
\ref{fig:theta_Seidov_SDEM}.

It is worthwhile to note that the major distinction between our
present work and the DEM calculations by
\citet{Bender_delta_expansion} and
 \citet{Seidov_LEE} is the
introduction of the simultaneous scale transformations shown in
equations \eqref{eqn:tt} and \eqref{eqn:xi_S_pi}. Through such a
scale transformation we have essentially performed a resummation
of the original series obtained by the DEM, directly leading to
the high accuracy of the SDEM. In fact, it can be shown that the
results derived by the DEM  can be recovered through (i) expanding
the SDEM results in powers of $n-1$ (or $n$), respectively; and
(ii)  truncating the resulting series at suitable orders.

\section{Mass of polytropes} \label{sec:approximants_mass}
Here we set to determine the local approximation of the mass of a
polytrope, $M$, as a function of the polytropic index $n$ by SDEM
analyses about $n=0$ or $n=1$. Furthermore, in Section
\ref{sec:global_approximants}, we will use these local
approximations to determine a global approximate expression for $M$
that is valid and accurate throughout the entire
interval $n\in[0,5]$.

The mass of a polytrope is given by:
\begin{equation}
\label{eqn:mass_polytropes_definition} M = \left[
\frac{(n+1)K}{4\pi G}\right]^{3/2} \rho_{c}^{(3-n)/(2n)} m(n),
\end{equation}
where the dimensionless mass function $m(n)$ and the second moment
of $\Theta^{n}$, $\mu(n)$, are respectively defined by:
\begin{align}
\label{eqn:mass_S} & m(n) = \frac{4}{\pi^2} \hat{\xi}^3 \mu_{}(n), \\
\label{eqn:mu2_def} &\mu_{}(n) = \int_{0}^{\pi}[ \Theta(z)]^{n}
z^2 dz.
\end{align}
As both $\hat{\xi}$ and $\Theta(z)$ have been obtained
perturbatively through the SDEM in the previous section, $m(n)$
can be found accordingly. Besides, it is interesting to note from
equation \eqref{eqn:mass_polytropes_definition} that the mass $M$
increases (decreases) with an increment in $\rho_{c}$ for $n<3$
($n>3$).  As the mass $M$ should be an increasing function of the
central density $\rho_{c}$ for stable stars with a given equation
of state, a polytropic star is thus stable (unstable) if the
polytropic index is less (greater) than 3, which is in agreement
with the standard result obtained from stability analysis
\citep[see, e.g.,][]{BWN}.

Using the SDEM results at $n=0$ or $1$, we form the local
approximant of the dimensionless mass $m(n)$, which is denoted by
$[m_{p}]^{k}_{j}(n)$  (with $p=0$ or $1$ signifying the perturbation
centre):
\begin{align}
[m_{p}]^{k}_{j}(n) %& =  4 \pi \left( S_{p}^{(0)} + \ldots + (n-p)^{j} S_{p}^{(j)} \right) \left(\mu_{p}^{(0)} + \ldots + (n-p)^{k} \mu_{p}^{(k)} \right) \nonumber \\
& = \frac{4}{\pi^2} \left\{ [\hat{\xi}_{p}]_{j}(n) \right\}^3
[\mu_{p}]_{k}(n) \label{eqn:m_local},
\end{align}
where the approximant of the normalised radius,
$[\hat{\xi}_{p}]_{j}(n)$, is defined in the previous section,
and $[\mu_{p}]_{k}$ is similarly defined:
\begin{align}
[\mu_{p}]_{k}(n) & = \mu_{p}^{(0)} + \ldots + (n-p)^{k}
\mu_{p}^{(k)} \label{eqn:m_local2},
\end{align}
and the subscript $j$ (superscript $k$)  of the notation
$[m_{p}]^{k}_{j}(n)$ denotes the perturbation order of the
normalized radius $\hat{\xi}(n)$ ($\mu(n)$).

We first consider the case at $p=0$. We note here that in
general $\mu_{0}^{(i)}$ is completely specified by
$\Theta_0^{(0)}, \Theta_0^{(1)},\ldots,\Theta_0^{(i-1)}$.
Similarly, we expand the dimensionless density term
$[\Theta(z)^{n}]$ in powers of $n$:
\begin{align}
\nonumber \Theta^{n} = & 1 +  n \ln \Theta_{0}^{(0)} + n^2 \left(\frac{1}{2} \ln ^2 \Theta_{0}^{(0)} + \frac{\Theta_{0}^{(1)}}{\Theta_{0}^{(0)}} \right) + n^3 \bigg(\frac{1}{6} \ln ^3 \Theta_{0}^{(0)} \\
& + \frac{\Theta_{0}^{(1)} \ln  \Theta_{0}^{(0)}}{\Theta_{0}^{(0)}}- \frac{\Theta_{0}^{(1)}{}^2}{2 \Theta_{0}^{(0)}{}^2}+\frac{\Theta_{0}^{(2)}}{\Theta_{0}^{(0)}}\bigg) + O[n^4].
\end{align}
By integrating order by order in $n$, it follows directly from the
expressions of $\Theta_0^{(0)}$, $\Theta_0^{(1)}$ and $\Theta_0^{(2)}$
obtained in Section \ref{sec:homology_perturbation}, we determine:
\begin{align}
\label{eqn:mu200_solution}
\mu_{0}^{(0)}& = \frac{\pi^3}{3} \approx 10.3354255601,\\
\mu_{0}^{(1)} & = \frac{\pi^3}{9} \left( -4 + 3 \ln 2 \right) \approx -13.2331926532, \label{eqn:mu201_solution} \\
\nonumber \mu_{0}^{(2)} & = \frac{\pi^3}{54}  \left(-21 \pi ^2+200+12 \ln 2 +36 \ln ^2 2\right) \\
& \approx 10.5377600301, \label{eqn:mu202_solution} \\
\nonumber \mu_{0}^{(3)} & = \frac{\pi^3}{81} \bigg(-3464+75 \pi ^2-210 \ln 2-63 \pi ^2 \ln 2\\
\nonumber & \quad +180 \ln ^2 2+36 \ln ^3 2+2646 \zeta (3)\bigg) \\
& \approx -8.12140718515. \label{eqn:mu203_solution}
\end{align}

From the SDEM calculations at $n=0$, we observe that the
approximant $[\mu_0]_k(n)$ appears to form a slowly converging
alternating series with a small radius of convergence in $n$.  To
accelerate the convergence of the series, we apply the Pad\'{e}
resummation technique to rewrite the approximant $[\mu_0]_k(n)$
 as a rational function, while respecting the series
expansion of the original series. In our case, we employ a $[2,1]$
Pad\'{e} approximant, $[\mu_0]_{[2,1]}(n)$, given explicitly by:
\begin{align}
[\mu_0]_{[2,1]}(n)= & \frac{1}{\left(\mu_{0}^{(2)} + n
\mu_{0}^{(3)}\right)}
\bigg\{ \mu_{0}^{(0)} \mu_{0}^{(2)} + n \bigg( \mu_{0}^{(1)}\mu_{0}^{(2)}  \nonumber \\
& - \mu_{0}^{(0)} \mu_{0}^{(3)} \bigg) + n^2
\left[(\mu_{0}^{(2)})^2 - \mu_{0}^{(1)} \mu_{0}^{(3)} \right]
\bigg\}, \label{eqn:mlocalhomology0P2}
\end{align}
to match $[\mu_0]_3(n)$
 up to the third order \citep[see, e.g.,][for the
construction and the notation of Pad\'{e}
approximants]{baker1975essentials}, in the hope that the
alternating behaviour is mimicked by the linear term in the
denominator.

\begin{figure}
    \centering
    \includegraphics[width = 0.45 \textwidth]{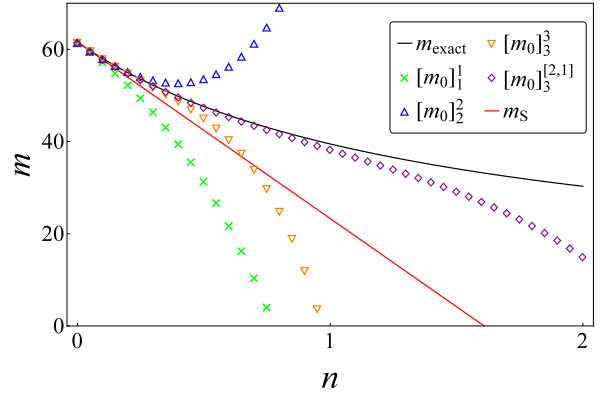}
    \caption{\label{fig:mass_local}
    The value of the mass $m$ is plotted as a function of the polytropic index $n$
    for the numerically exact solution (black line);
    the diagonal SDEM approximants about  $n=0$, $[m_{0}]_{j}^{j}$ in equation (\ref{eqn:m_local})
    for $j=1$ (green cross), $j=2$ (blue triangle), and $j=3$ (orange
    triangle);
    the $[2,1]$ Pad\'{e} approximant, $[m_{0}]^{[2,1]}_{3} (n)$
    in equation (\ref{eqn:mlocalhomology0P}) (purple rhombus);
    and the first order DEM result about $n=0$ (see equation (\ref{eqn:Seidov_mass}), red line).
    }
\end{figure}

In Figure \ref{fig:mass_local}, we compare the accuracy of several
local approximants of  $m$ about $n=0$, including
$[m_{0}]^{1}_{1}$, $[m_{0}]^{2}_{2}$, $[m_{0}]^{3}_{3}$, and the
Pad\'{e} approximant:
\begin{align}
[m_{0}]^{[2,1]}_{3} (n)= & \frac{4}{\pi^2}\left\{
[\hat{\xi}_{0}]_{3}(n) \right\}^3 [\mu_0]_{[2,1]}(n),
\label{eqn:mlocalhomology0P}
\end{align}
where the third order approximant of the normalized radius,
$[\hat{\xi}_{0}]_{3}(n)$, is adopted specifically for illustration
purpose. Numerical results reveal that, except for the Pad\'{e}
approximant $[m_{0}]^{[2,1]}_{3} (n)$, the other approximants are
only accurate near the perturbation center $n=0$ and have a narrow
interval of validity. The errors build up rapidly when $n$ is
close to unity. In contrast, the $[2,1] $ Pad\'{e} approximant
$[m_{0}]^{[2,1]}_{3} (n)$ closely resembles the numerical solution
over an extended interval from $n=0$ to $n \approx 1.5$. It
confirms the necessity of an appropriate series resummation.

We also show in Figure \ref{fig:mass_local}  the approximant of
the mass derived   from the DEM \citep{Seidov_LEE}, which is given
by:
\begin{equation}
\label{eqn:Seidov_mass} m_{\text{S}}(n) = 8 \sqrt{6} \pi +\frac{ 4
\sqrt{6} \pi }{3} (-37+48 \ln 2) n + O[n^2].
\end{equation}
The performance of $m_{\text{S}}(n)$ is quite close to that of
other direct SDEM expansions such as $[m_{0}]^{1}_{1}$,
$[m_{0}]^{2}_{2}$, $[m_{0}]^{3}_{3}$. It is accurate only within a
narrow range extending from  $n=0$ to $n \approx 0.5$. Once again
it pinpoints the importance of the application of the Pad\'{e}
approximation in order to enlarge the domain of validity of the
approximant of the mass.

Here we put forward an argument to support the necessity of
introducing the Pad\'{e} approximant to resum the expansion of
$\mu(n)$.
We notice that, to the leading order, the second moment integral
$\mu(n)$ is given by:
\begin{equation}
\mu(n) = \int_{0}^{\pi} z^2 (1 - \frac{z^2}{\pi^2} )^n dz = \frac{\pi ^{7/2} \Gamma (n+1)}{4 \Gamma (n+ 5/2)},\quad \mbox{for} \quad n>-1, \label{eqn:mass_leading}
\end{equation}
where $\Gamma(z)$ is the gamma function defined by \citep[see,
e.g.,][]{Math_handbook}:
\begin{equation}
\Gamma(z) = \int_{0}^{\infty} t^{z-1} e^{-t} dt.
\end{equation}
%When $n\leq -1$, $\mu(n)$ in ill-defined in real for the diverging surface contribution.
It is a well-known fact that $\Gamma(n+1)$ is a meromorphic function with
simple poles at $n=-1,-2,-3, \ldots$. In the SDEM calculations
about $n=0$, we have expanded the $\Theta^n$ term into a power
series of $n$, and the $n$-series expansion has the radius of
convergence $1$ due to the nearest singularity at $n=-1$. It then
readily  explains the narrow interval of validity of the direct
expansions  of $\mu(n)$ about the point $n=0$.
On the other hand, numerical investigation suggests that the
integral for $\mu(n)$ (see equation \eqref{eqn:mu2_def}) can be
computed more accurately if $[\Theta(z)]^n$ there is expanded through
binomial expansion instead of the SDEM. However, the
resultant integrals do not have simple solutions in terms of
elementary mathematical functions.

On the other hand, we can similarly apply the SDEM to expand the
integral $\mu(n)$ about $n=1$ to form a power series of $n-1$ (see
equation \eqref{eqn:m_local2}), with the leading two expansion
coefficients given explicitly by:
\begin{align}
\mu_{1}^{(0)} & = \pi, \label{eqn:mu210_solution} \\
\nonumber \mu_{1}^{(1)} & = \frac{\pi}{4} [\text{Cin}(2 \pi ) + 6 - 4 \ln(2\pi) ]-\frac{1}{2} \text{Si}(2 \pi) \\
& \approx -3.68507639562. \label{eqn:mu211_solution}
\end{align}
However, higher order coefficients are not shown 
due to their cumbersome expressions. Although we could still find
accurate local approximant of $m(n)$ from $\mu_{1}^{(0)}$ and
$\mu_{1}^{(1)}$, its applicability is limited within a narrow
region where $n \approx 1$. Instead of further expanding
$\mu(n)$ about $n=1$,  we shall combine all the information
resulting from expansion about $n=0$ and $n=1$  in Section
\ref{sec:global_approximants} using the two-point Pad\'{e}
approximation technique to enhance the accuracy and extend the interval of
validity of the relevant approximant.

\section{Global Approximation of Polytropes} \label{sec:global_approximants}
So far we have determined the variations of the polytrope
function, the scale factor and the mass up to the third
order about $n=0$, as well as such variations up to the first
order about $n=1$ based on the SDEM.  These results constitute
local approximations about $n=0$ and $n=1$. On the other hand, in
addition to the exact solution of the polytrope function at $n=5$
(see equation \eqref{eqn:exact5}), \citet{Buchdahl_n5} obtained
the following leading order approximation of $\hat{\xi}(n)$:
\begin{equation}
\hat{\xi}(n) = \frac{32 \sqrt{3}}{\pi (5- n)} + O[1],
\label{eqn:n5}
\end{equation}
which is valid for polytropes with $n$ less than and close to 5
(see Appendix \ref{sec:literature_review23} for a brief account of
Buchdahl's argument).
%Besides, we also find that this
%approximation can  be derived by performing DEM about $n=5$ and
%the details are given in Appendix \ref{sec:literature_review23}.
In
this section, starting from the perturbation results as well as
the analytically exact solution about $n=0,1$ and $5$, we derive
globally valid, approximate expressions for the normalised radius, the
the mass and the polytrope function and as a function of the
polytropic index $n$. The spirit is to construct expressions that
respect the local variations about $n=0$ and $n=1$, and also match
the exact value or the asymptotic behaviour about $n=5$. These
global approximants will be compared to other existing
approximants reviewed in Section \ref{sec:literature_review1} and
Appendix \ref{sec:literature_review23}.  For ease of reference,
the useful results are presented in a self-contained manner in
Appendix \ref{sec:formulae_for_lazy_physicists}.

\subsection{Normalised radius}
As the normalised radius of a polytrope is directly proportional to
$S(n)^{(n-1)/2}$ (see equation \eqref{eqn:xi_S_pi}), we first look
for a global approximant of the scale factor $S$. A modified
$[3,2]$ Pad\'{e} approximant of  $S(n)$ is proposed:
\begin{equation}
\label{eqn:homology_S_approximant}
S_{\text{g}}(n) = \frac{a_{1} + a_{2} n + a_{3} n^2 + a_{4} n^3}{ \sqrt{5-n} \big(1 + a_{5} n +  a_{6} n^2 \big)},
\end{equation}
where the constants $a_{1}$, ..., $a_{6}$ are determined by the
SDEM results of $S(n)$ about $n=0$ and $n=1$, and the square root
term in the denominator is chosen to match the leading behaviour
of the radius when $n$ approaches $5$ (see equation \eqref{eqn:n5}
and Appendix \ref{sec:literature_review23}).

Using the perturbation coefficients $S_{0}^{(0)}$, $S_{0}^{(1)}$,
$S_{0}^{(2)}$, $S_{0}^{(3)}$,  $S_{1}^{(0)}$ and $S_{1}^{(1)}$,
whose values are given  in equations (\ref{eqn:S00_solution}),
(\ref{eqn:S01_solution}), (\ref{eqn:S02_solution}),
(\ref{eqn:S03_solution}), (\ref{eqn:S10_solution}) and
(\ref{eqn:S11_solution}), respectively, we can fix the values of
$a_{1}, \ldots, a_{6}$:
\begin{equation*}
\begin{split}
& a_{1} \approx 3.678184391977817, \\
& a_{2} \approx  -0.12127837785202653, \\
& a_{3} \approx -0.0820898766826553,\\
& a_{4} \approx 0.0030327766768460046, \\
& a_{5} \approx 0.00858273787249898,\\
& a_{6} \approx -0.018845815183087977.
\end{split}
\end{equation*}
Moreover, we have checked that
the Pad\'{e} approximant in equation
\eqref{eqn:homology_S_approximant} does not diverge within the
physical range $n \in[0,5)$.

When $S_{\text{g}}(n)$ is substituted into equation
\eqref{eqn:xi_S_pi}, accurate values of the normalised radius can readily be
reproduced. However, to further improve the accuracy, we propose
the following semi-analytical approximant of the normalised radius:
\begin{equation}
\label{eqn:xi_homology} \hat{\xi}_{\text{g}}(n) = \frac{ \pi
S_{\text{g}}(n)^{(n-1)/2}}{1 + a_{0} n^{8}(n-1)^{8}} .
\end{equation}
Here the constant $a_{0}$ is fixed by imposing the leading
bahaviour at $n=5$ in equation (\ref{eqn:n5}), which shows that as
$n \rightarrow 5$ from the left,  $\hat{\xi}(n) \rightarrow (32
\sqrt{3})/[ \pi (5- n)]$. While the simple pole-like divergence is
taken care of in the scale factor, the constant $a_{0}\approx
1.5996644405401317 \times 10^{-17}$ now accounts for the
residue-like factor $32 \sqrt{3}/\pi$. Besides, the high power
$n^{8}(n-1)^{8}$ is added to minimize the influence of the
denominator on the local approximations around $n=0$ and $n=1$.

\begin{figure}
    \includegraphics[width = 0.49 \textwidth]{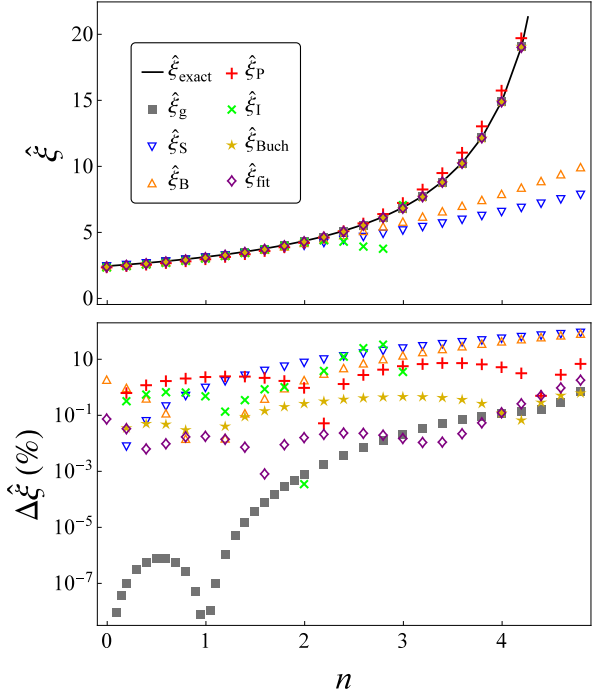}
    \caption{\label{fig:first_zero}
    In the upper panel, the value of the normalised radius $\hat{\xi}$ is plotted
    as a function of the polytropic index $n$ for the numerically
    exact solution $\hat{\xi}_{\text{exact}}$ (black line),
    $\hat{\xi}_{\text{g}}$ in equation (\ref{eqn:xi_homology})  (grey square),
    $\hat{\xi}_{\text{S}}$ in equation (\ref{eqn:Seidov_xi}) (blue triangle),
    $\hat{\xi}_{\text{B}}$ in equation (\ref{eqn:Bender_xi}) (orange triangle),
    $\hat{\xi}_{\text{P}}$ (see Appendix \ref{sec:literature_review23} and equation
   (\ref{eqn:Pascual_theta}), red plus),
    $\hat{\xi}_{\text{I}}$ in equation (\ref{eqn:Iacono_xi}) (green cross),
    $\hat{\xi}_{\text{Buch}}$
    in equation (\ref{eqn:Buchdahl_xi}) (yellow star), and
     $\hat{\xi}_{\text{fit}}$
    in equation (\ref{eqn:Iacono_fit_xi}) (purple rhombus).
    The associated percentage error plot in logarithmic scale is shown in the lower panel,
    with $\Delta \hat{\xi} = |\hat{\xi} - \hat{\xi}_{\text{exact}}|/\hat{\xi}_{\text{exact}} \times 100\%$.}
\end{figure}

We show the global approximant of the normalised radius,
$\hat{\xi}_{\text{g}}(n)$ in equation \eqref{eqn:xi_homology}, as
a function of $n$ in Figure \ref{fig:first_zero}, where the exact
numerical value of the normalised radius, $\hat{\xi}_{\text{exact}}(n)$, and
other approximants reviewed in the text and Appendix
\ref{sec:literature_review23} (including
$\hat{\xi}_{\text{S}}(n)$, $\hat{\xi}_{\text{B}}(n)$,
$\hat{\xi}_{\text{P}}(n)$ $\hat{\xi}_{\text{I}}(n)$,
$\hat{\xi}_{\text{Buch}}(n)$ and $\hat{\xi}_{\text{fit}}(n)$) are
shown as well. First of all, we see that the normalised radius
increases with the polytropic index $n$. It agrees with
our intuition that the density tail of a polytrope extends with
increasing polytropic index $n$. Secondly, as for the accuracy of
various approximation schemes (see the lower panel of Figure
\ref{fig:first_zero}), we find that $\hat{\xi}_{\text{g}}(n)$ is
better than other existing approximations. While the approximants
derived from the DEM \citep{Seidov_LEE,Bender_delta_expansion},
 $\hat{\xi}_{\text{S}}(n)$ in equation
(\ref{eqn:Seidov_xi}) and $\hat{\xi}_{\text{B}}(n)$ in equation
(\ref{eqn:Bender_xi}), are only accurate in the vicinity of their
respective perturbation centers, the error of
$\hat{\xi}_{\text{g}}(n)$  stays below $1$ per cent throughout the
physical range $n \in [0,5)$. In fact,  it is less than $8.1
\times 10^{-7}$ per cent for $n\in[0,1]$. In comparison, over the
same range $n\in[0,1]$, the maximum error of
$\hat{\xi}_{\text{P}}(n)$ in equation (\ref{eqn:Pascual_theta})
\citep{Pascual_Pade},
 $\hat{\xi}_{\text{Buch}}(n)$ in equation
(\ref{eqn:Buchdahl_xi}) \citep{Buchdahl_n5} and
$\hat{\xi}_{\text{fit}}(n)$ in equation (\ref{eqn:Iacono_fit_xi})
\citep{Iacono} reach $2.6$, $0.055$ and $0.079$ per cent,
respectively. Besides, the Pad\'{e} approximant
$\hat{\xi}_{\text{I}}(n)$ in equation (\ref{eqn:Iacono_xi})
\citep{Iacono}  diverges at $n = 3.050$, and is ill-defined in
real beyond that point.

\subsection{Mass}
We construct an approximant that matches the SDEM results about
$n=0$ and $n=1$, as well as the analytically exact value of the
mass $m$ at $n=5$. It follows directly from   \eqref{eqn:exact5}
that:
\begin{equation}
\label{eqn:mass_n5}
m (n=5) = 4 \pi \int_{0}^{\infty} \frac{x^2}{(1 + x^2/3)^{5/2}} dx = 4 \pi \sqrt{3}.
\end{equation}
As shown in equation \eqref{eqn:homology_S_approximant},
$\hat{\xi}_{\rm g}$ diverges as $(5-n)^{-1}$ as $n$ approaches 5
from the left (see equations \eqref{eqn:homology_S_approximant}
and \eqref{eqn:xi_homology}). For $m$ to remain finite and
non-vanishing at $n=5$, we hypothesize an approximant of the
second moment integral $\mu_{\rm g}(n)$ to carry a factor of
$(5-n)^3$ in the form:
\begin{equation}
\label{eqn:mu2h_approximant} \mu_{\rm g}(n) = (5-n)^3 \frac{b_{1}
+ n b_{2} + n^2 b_{3} + n^3 b_{4}}{1 + n b_{5} + n^2 b_{6}},
\end{equation}
where the real constants $b_{i}$ for $i=1,2,\ldots,6$ are to be
determined by the SDEM results at $n=0$ (i.e., $\mu_{0}^{(0)}$,
$\mu_{0}^{(1)}$, $\mu_{0}^{(2)}$, and $\mu_{0}^{(3)}$)
%(i.e., $\mu_{2,0}^{(0)}$, $\mu_{2,0}^{(1)}$, $\mu_{2,0}^{(2)}$,
%and $\mu_{2,0}^{(3)}$)
 and $n=1$ (i.e., $\mu_{1}^{(0)}$ and $\mu_{1}^{(1)}$).
%(i.e., $\mu_{2,1}^{(0)}$ and $\mu_{2,1}^{(1)}$).
The values of
these constants can be readily found from  equations
(\ref{eqn:mu200_solution}), (\ref{eqn:mu201_solution}),
(\ref{eqn:mu202_solution}), (\ref{eqn:mu203_solution}),
(\ref{eqn:mu210_solution}) and (\ref{eqn:mu211_solution}), which
are given by:
\begin{equation*}
\begin{split}
& b_{1} \approx 0.08268340448079952, \\
& b_{2} \approx  0.0570923774427696, \\
& b_{3} \approx -0.0021371524111317, \\
& b_{4} \approx -0.000863277094516044, \\
& b_{5} \approx 1.370866096910041, \\
& b_{6} \approx 0.415498502167336.
\end{split}
\end{equation*}

We check that the denominator vanishes at $n \approx -2.21$ and $n
\approx -1.09$, so the Pad\'{e} approximant does not diverge
within the physical range $n \in[0,5]$. It echoes the leading
behaviour of the mass in equation
(\ref{eqn:mass_leading}), that the $\Gamma(n+1)$ factor contains
simple poles at $n=-1, -2, -3, \ldots$.

So far, we have imposed the SDEM results at $n=0$ and at $n=1$.
To utilize the analytically exact value of $m(n=5)$, we propose an
approximant in the form:
\begin{equation}
\label{eqn:mass_homology_soln} m_{\text{g}}(n) =4 \pi
S_{\text{g}}(n)^{3(n-1)/2} \mu_{\rm g}(n)+ b_{0} (5-n)^{(15-3n)/4}
n^8 (n-1)^8  ,
\end{equation}
While the first term in the RHS of the above equation follows
directly from equations \eqref{eqn:xi_S_pi} and
\eqref{eqn:mass_S}, the second term is a correction term aimed at
reproducing the exact mass  at $n=5$ (see equation
(\ref{eqn:mass_n5})). To this end, the real constant $b_{0}
\approx -3.420867516502784\times 10^{-10}$ is chosen and the
factor $(5-n)$ is raised to the power $(15-3n)/4$ so that the
correction term matches the behaviour of $S_{\text{g}}^{3(n-1)/2}
\mu_{\rm g}$ as $n$ approaches $5$ from the left. Besides,  we
have multiplied the correction term by a factor $n^8 (n-1)^8$ to
minimize the influence of the correction term on $m_{\text{g}}(n)$
for $n$ close to 0 or 1.

\begin{figure}
    \includegraphics[width = 0.49 \textwidth]{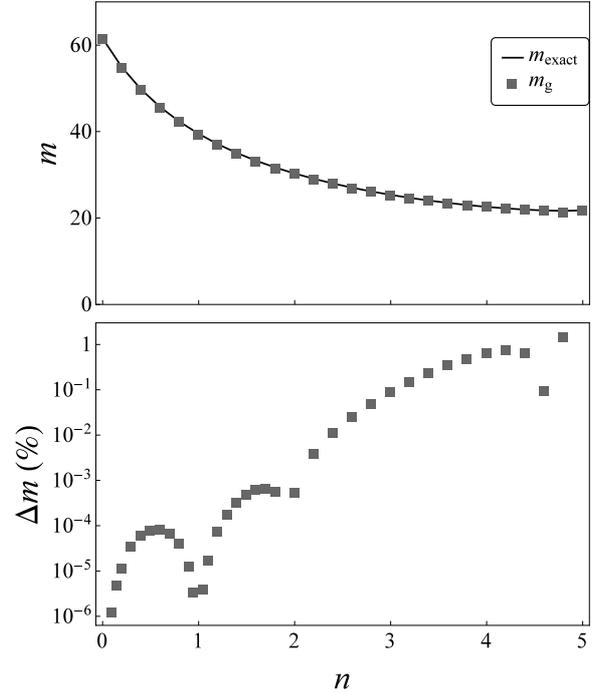}
    \caption{\label{fig:mass}
       In the upper panel, the value of the dimensionless mass $m$ is plotted as a function
       of $n$ for the numerically exact solution (black line), and
    the global SDEM approximant $m_{\text{g}}$ in equation (\ref{eqn:mass_homology_soln})  (grey square),
    The associated error plot in logarithmic scale is shown in the lower panel,
    with $\Delta m = |m - m_{\text{exact}}|/m_{\text{exact}} \times 100\%$.}
\end{figure}

The numerical solution and the global approximant of the
mass, $m_{\text{g}}(n)$, are plotted in Figure
\ref{fig:mass}.  We see that $m_{\text{g}}(n)$ can nicely
approximate the numerical value with a maximum error of about
$2\%$  in the whole range where $n \in [0,5]$. Such high degree of
accuracy is achieved by  incorporating the SDEM results obtained
at $n=0$ and $n=1$ into a Pad\'{e} approximant, together with a
correction term such that $m_{\text{g}}(n)$ agrees with the
analytically exact value at $n=5$. It is worthwhile to remark that
such a global approximant of the mass is derived
analytically from the SDEM without the help of any numerical
fittings.

On the other hand, we can also see the dependence of the mass $m$
as a function of $n$ in Figure \ref{fig:mass}. While the
polytropic tail lengthens with increasing polytropic index, the
density decreases in the scaled frame $z$. Overall, the decrease
of the density term dominates the lengthening of the polytropic
tail, so that the total dimensionless mass decreases with
increasing polytropic index $n$. Notice that the lengthening of
the polytropic tail is less pronounced near the incompressible
limit $n=0$, as could be seen in Figure \ref{fig:first_zero}. As a
result, the mass decreases more rapidly
near the incompressible limit $n=0$.

\subsection{Normalised polytrope function}

Based on the SDEM results at $n=0$ and $n=1$, we construct a two-point polynomial approximation
for the polytrope function $\hat{\theta}_{\text{tp}}(x)$,
by requiring its variations to match the local SDEM
results about $n=0$ and $n=1$.  The approximant is given by:
\begin{align}
\label{eqn:tp_theta}
\nonumber \hat{\theta}_{\text{tp}}(x) =&  \Theta_0^{(0)}(z_{\text{g}})+n \Theta_0^{(1)}(z_{\text{g}})+n^2 \Theta_0^{(2)}(z_{\text{g}}) +n^3 \bigg[-4 \Theta_0^{(0)}(z_{\text{g}}) \\
&\nonumber -3 \Theta_0^{(1)}(z_{\text{g}}) -2 \Theta_0^{(2)}(z_{\text{g}}) + 4 \Theta_1^{(0)}(z_{\text{g}})-\Theta_1^{(1)}(z_{\text{g}})\bigg] \\
&\nonumber +n^4 \bigg[3 \Theta_0^{(0)}(z_{\text{g}}) +2 \Theta_0^{(1)}(z_{\text{g}})+\Theta_0^{(2)}(z_{\text{g}})\\
& -3
\Theta_1^{(0)}(z_{\text{g}})+\Theta_1^{(1)}(z_{\text{g}})\bigg],
\end{align}
where the scaled length scale $z_{\text{g}} = \pi
x/\hat{\xi}_{\text{g}}(n)$.

\begin{figure}
    \includegraphics[width = 0.49 \textwidth]{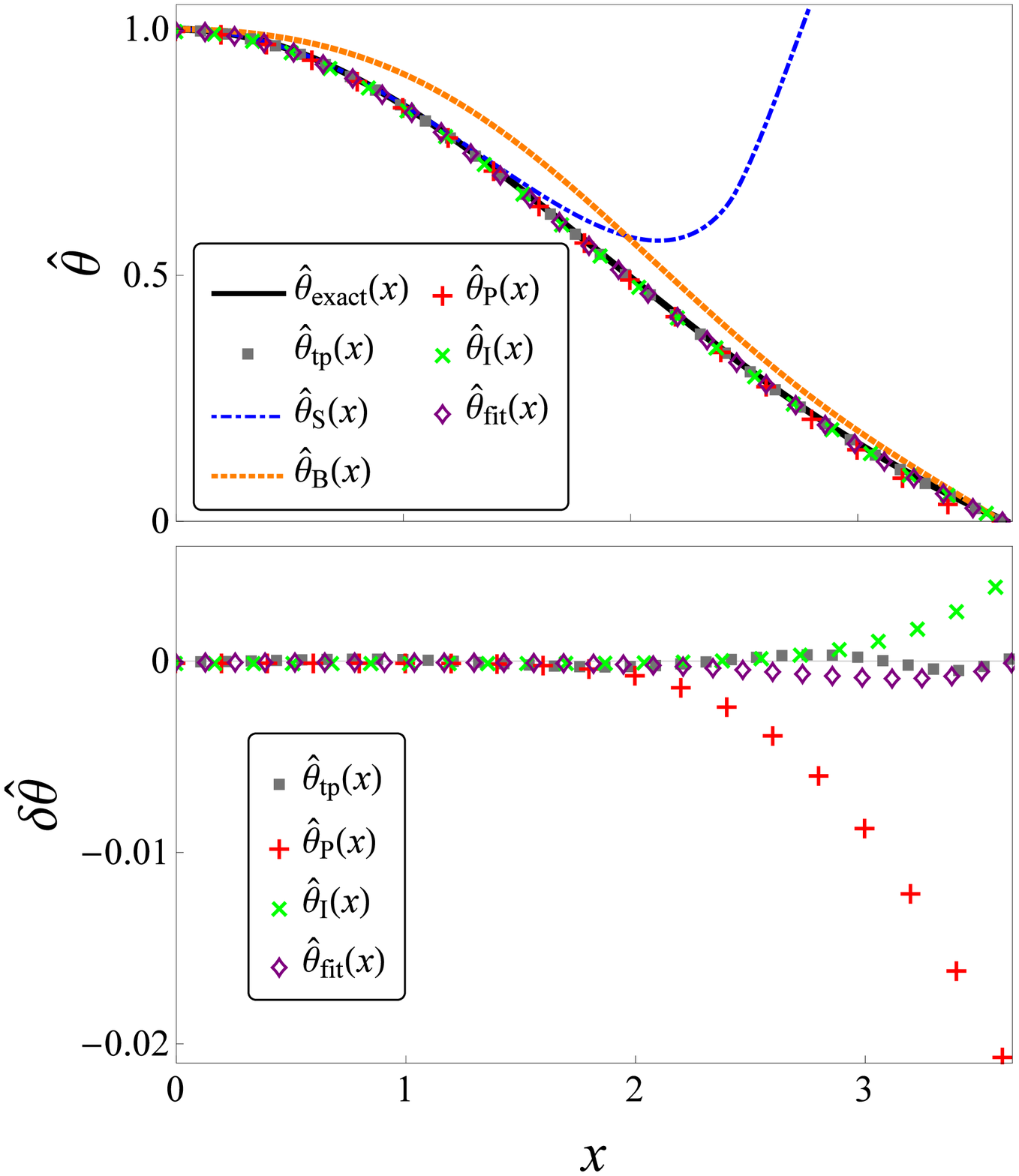}
    \caption{\label{fig:n15_theta}
    In the upper panel, for $n=1.5$, we plot
    the numerically exact polytrope function $\hat{\theta}_{\text{exact}}(x)$ (black solid line),
    the two-point SDEM approximant $\hat{\theta}_{\text{tp}}(x)$ in equation (\ref{eqn:tp_theta}) (grey square),
     $\hat{\theta}_{\text{S}}(x)$ \citep{Seidov_LEE} in equation (\ref{eqn:Seidov_theta}) (blue dot-dashed),
     $\hat{\theta}_{\text{B}}(x)$ \citep{Bender_delta_expansion}  in equation (\ref{eqn:Bender_theta}) (orange dotted),
     $\hat{\theta}_{\text{P}}(x)$ \citep{Pascual_Pade} in equation (\ref{eqn:Pascual_theta}) (red
     plus),
     $\hat{\theta}_{\text{I}}(x)$ \citep{Iacono} in equation (\ref{eqn:Iacono_theta}) (green cross), and
     $\hat{\theta}_{\text{fit}}(x)$ \citep{Iacono} (see equation (\ref{eqn:Iacono_fit_xi})) (purple rhombus)
    Plotted in the lower panel is the associated deviation from the exact polytrope function,
    $\delta \hat{\theta} = \hat{\theta} - \hat{\theta}_{\text{exact}}$ for
    the approximants, $\hat{\theta}_{\text{g}}(x)$, $\hat{\theta}_{\text{P}}(x)$,
    $\hat{\theta}_{\text{I}}(x)$, and $\hat{\theta}_{\text{fit}}(x)$.
    }
\end{figure}

Next, in the upper panel of Figure \ref{fig:n15_theta} we compare
the two-point SDEM approximant of the polytrope function,
$\hat{\theta}_{\text{tp}}(x)$ in equation
(\ref{eqn:tp_theta}), against other existing approximations,
including the two formulae obtained from the DEM, i.e.,
$\hat{\theta}_{\text{S}}(x)$ in equation (\ref{eqn:Seidov_theta})
\citep{Seidov_LEE} and $\hat{\theta}_{\text{B}}(x)$ in equation
(\ref{eqn:Bender_theta}) \citep{Bender_delta_expansion}, and
 the two Pad\'{e} approximants
$\hat{\theta}_{\text{P}}(x)$ \citep[see \eqref{eqn:Pascual_theta}
and][]{Pascual_Pade},
$\hat{\theta}_{\text{I}}(x)$ and $\hat{\theta}_{\text{fit}}(x)$ \citep[see
\eqref{eqn:Iacono_theta} and][]{Iacono}, for the case $n=1.5$.
Several remarks about the performance of various approximants are
in order. First, we note that $\hat{\theta}_{\text{tp}}(x)$ is
indeed very close to the exact value
$\hat{\theta}_{\text{exact}}(x)$ and their difference is smaller
than 0.001 throughout the polytrope.  Secondly, as mentioned
previously, $\hat{\theta}_{\text{S}}(x)$ and
$\hat{\theta}_{\text{B}}(x)$ are ill-defined in real for $x >
\sqrt{6}$ and $x>\pi$, respectively, because they fail to capture
the movement of the branch point singularity at the surface of
the polytrope. We take the real part of
$\hat{\theta}_{\text{S}}(x)$ and $\hat{\theta}_{\text{B}}(x)$ in this
figure. Still, $\hat{\theta}_{\text{S}}(x)$  deviates significantly
from the exact value near the surface of the polytrope. On the
other hand, while $\hat{\theta}_{\text{B}}(x)$ can nicely approximate
$\hat{\theta}_{\text{exact}}(x)$ at the centre and the surface of
the polytrope, its accuracy worsens in the intermediate range.

As for case of the two analytically derived Pad\'{e} approximants
$\hat{\theta}_{\text{P}}(x)$  and $\hat{\theta}_{\text{I}}(x)$,
which are obtained by resumming the series solution of the LEE
about $x=0$, we see that they are highly accurate near $x=0$ but
the error builds up all the way to the surface of the polytrope.
In comparison, the two-point SDEM approximant
$\hat{\theta}_{\text{tp}}(x)$ oscillates about the exact value
with a tiny amplitude and is uniformly accurate throughout the
entire physical region.  The accuracy of the two-point SDEM
approximant  $\hat{\theta}_{\text{tp}}(x)$ is   similar to that of
the approximant $\hat{\theta}_{\text{fit}}(x)$, which are the only
two approximations that are uniform throughout the physical domain
of interest. A good approximation of the polytrope function near
the surface is associated with an accurate determination of the
radius. In the two-point SDEM analysis, the radius is accurately
and analytically determined via a scale transformation, and could
be systematically improved by increasing the perturbation order.
The latter relies on least square fitting against numerical
results, and its accuracy could be boosted by using more free
parameters in the fitting.

As the normalised radius of the polytrope at $n=5$ is infinite,
SDEM is not directly applicable at $n=5$.  On other other hand,
$\hat{\theta}_{\text{P}}(x)$ in equation (\ref{eqn:Pascual_theta})
utilises the exact solution at $n=5$ by an ingenious variable transformation.
For an analytical approximant that is accurate for all finite polytropes
over the range of $n$ in $[0,5)$, we define an piecewise approximant to be
\begin{equation}
\label{eqn:theta_global}
\hat{\theta}_{\text{g}} (x) = \begin{cases}
\hat{\theta}_{\text{tp}}(x), & \text{for} \qquad 0 \leq n \leq 2, \\
\hat{\theta}_{\text{P}}(x), & \text{for} \qquad 2 < n < 5.
\end{cases}
\end{equation}

\begin{figure}
    \includegraphics[width = 0.49 \textwidth]{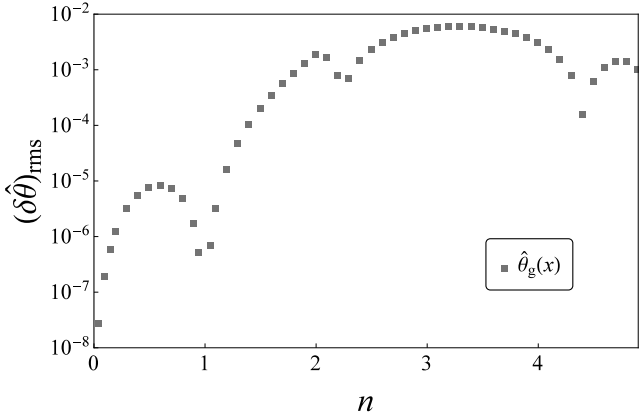}
    \caption{\label{fig:theta_global_error}
    The root mean square error $(\delta \hat{\theta})_{\text{rms}}
    \equiv [\int_{0}^{\hat{\xi}} [\hat{\theta}(x) - \hat{\theta}_{\text{exact}}(x)]^2 dx / \hat{\xi}]^{1/2} $
    of the global approximant $\hat{\theta}_{\text{g}}(x)$ in equation (\ref{eqn:theta_global}) (grey
    square) is plotted against the polytropic index $n$.
    }
\end{figure}

In Figure \ref{fig:theta_global_error}, we see that the root mean
square error $(\delta \hat{\theta})_{\text{rms}}
    \equiv [\int_{0}^{\hat{\xi}} [\hat{\theta}(x) - \hat{\theta}_{\text{exact}}(x)]^2 dx / \hat{\xi}]^{1/2} $
    of the approximant $\hat{\theta}_{\text{g}}(x)$ is less than $6.1 \times
10^{-3}$ throughout the physical range of finite polytropes for $n
 \in [0,5)$.  In particular, $(\delta \hat{\theta})_{\text{rms}}$
    is within $8.4 \times 10^{-6}$ between the two SDEM perturbation
centres at $n=0$ and $1$.  The error increases sharply from $n=1$
to $n=2$ signifying a decline of accuracy of the two-point SDEM
approximant  $\hat{\theta}_{\text{tp}}(x)$. Beyond $n=2$, the
Pad\'{e} approximant $\hat{\theta}_{\text{P}}(x)$ is more accurate
than the two-point SDEM approximant $\hat{\theta}_{\text{tp}}(x)$.
The Pad\'{e} approximant $\hat{\theta}_{\text{P}}(x)$ is therefore
used to define the global approximant of the normalised polytrope
function.

\section{Discussion and Conclusion} \label{sec:conclusion}

Motivated by the successful application of the DEM to the solution
of the LEE \citep{Bender_delta_expansion,Seidov_LEE}, we propose
in the present paper the SDEM, an extension of the DEM, to remedy
the inadequacy of the original scheme. The major problem with the
LEE is the $\theta^n$ term, which  in general ceases to be an
analytic function at points where $\theta$ vanishes. In fact, the
$\theta^n$ term could become complex-valued if $\theta$ is less
than zero. As a consequence, the DEM is expected to break down
outside the unperturbed polytrope (with $n=0$ or $1$).
As the normalised radius of a polytrope is an increasing function of the
polytropic index $n$, in the DEM the polytrope function of a
perturbed polytrope with $n>0$ (or $n>1$) usually becomes complex
near its surface. Even if the real part of such a complex function
is taken, the accuracy of the resultant polytrope function could
be poor (see Figures   \ref{fig:theta_Seidov_SDEM} and
\ref{fig:n15_theta}).

The SDEM is aimed at solving the above-mentioned problem. We
exploit the scale invariance symmetry of the LEE to define an
$n$-dependent scale transformation  under which the scaled
polytrope function $\Theta(z)$ is required to attain its first
zero at the scaled coordinate $z=[S(n)]^{(1-n)/2} x =\pi$
regardless of the value of $n$. As a result, the domain of
definition of the scaled LEE, where the polytrope function is real
and non-negative, remains fixed throughout the perturbation
scheme. In addition, the LEE is transformed from an initial value
problem to an eigenvalue problem with the scale factor $S(n)$
being the eigenvalue (see equation
\eqref{eqn:homology_perturbation_problem}).  In the present paper
we have derived the perturbation corrections about $n=0$ up to the
third order, and that about $n=1$ up to the first order. The local
SDEM approximants are compared to the DEM results, and the former
are found to be more accurate. In particular, the SDEM result is
accurate within the entire polytrope and consequently the
radius of a polytrope can also be determined accurately.

From the SDEM,  we have derived the local as well as the global
approximations to the normalised radius, the mass,
and the normalised polytrope
function as a function of the polytropic
index $n$. The global approximant of the normalised radius,
$\hat{\xi}_{\text{g}}(n)$ in equation (\ref{eqn:xi_homology}), has
a maximum percentage error of $8.1 \times 10^{-7}$ per cent ($1$
per cent)  for $n\in[0,1]$ ($n\in [0,5)$). Furthermore, the global
SDEM approximant of the mass carries the maximum percentage error
of $8.5 \times 10^{-5}$ per cent ($2$ per cent) for $n$ in $[0,1]$
($ n\in [0,5)$). All such approximants are derived analytically
without adoption of numerical fitting.

Two remarks about the scale transformation are in order. First,
the choice of $\pi$ as the radius in the new scale indeed
offers an additional benefit of an exact value of $\hat{\xi}(n=1)
= \pi$ (see equation (\ref{eqn:xi_S_pi})) irrespective of the
value of the  scale factor $S(n)$. Therefore, as far as
$\hat{\xi}$ is concerned, the SDEM expansion about $n=0$ is still
able to yield the exact value at $n=1$ (see Figure
\ref{fig:xi_Seidov_SDEM}). Secondly, the length dependence and the
density dependence are separated in the SDEM calculations. For
each physical quantity of a polytrope, each physical length factor
is associated with a $S^{(n-1)/2}$ factor, or equivalently a
$\hat{\xi}$ factor, while each density term is proportional to
$\Theta(z)^{n}$.

In principle we can extend the SDEM calculations about $n=0$ and
$n=1$ to high orders. The convergence property of such
perturbation series is an interesting issue. On the other hand, it
is also possible to apply the SDEM at other values of $n$ with no
known closed form solution by numerical means. These researches
are under way and relevant results will be reported elsewhere in
due course.

%%%%%%%%%%%%%%%%%%%%%%%%%%%%%%%%%%%%%%%%%%%%%%%%%%

\section*{Acknowledgements}
We thank C.F. Lo at the Physics Department, The Chinese University
of Hong Kong for fruitful discussion and valuable suggestions.



\bibliographystyle{mnras}
\bibliography{LEE_bibi_v5} % if your bibtex file is called example.bib

%%%%%%%%%%%%%%%%%%%%%%%%%%%%%%%%%%%%%%%%%%%%%%%%%%

%%%%%%%%%%%%%%%%% APPENDICES %%%%%%%%%%%%%%%%%%%%%

\appendix

\section{Other analytical approximations to the LEE} \label{sec:literature_review23}

While the application of the DEM to the standard LEE has been
reviewed in Section \ref{sec:literature_review1}, here we
summarize the results of some other relevant  approximation
schemes which have been used as the benchmarks for the accuracy of
the SDEM established in the present paper.

In order to obtain a uniform approximation of the polytrope
function and its normalised radius, based on the power series solution of
the standard LEE and also in the light of the analytical solution
of LEE at $n=5$, \citet{Pascual_Pade} introduced a scaled
coordinate $z_{\text{P}}$:
\begin{equation} \label{eqn:zP}
z_{\text{P}} = 6 \bigg[
\bigg( 1 + \frac{x^2}{3} \bigg)^{1/2} -1 \bigg]
\end{equation}
to replace the standard independent variable $x$ of the LEE. After
expressing the traditional power series solution of the LEE in a
power series in $z_{\text{P}}$ and and performing a Pad\'{e}
resummation of such a
 series, \citet{Pascual_Pade} obtained the Pad\'{e} approximants of various orders for the polytrope
 function. In particular, the [2,2] Pad\'{e} approximant
 \citep[see, e.g.,][]{baker1975essentials} of the polytrope function is given by:
%\begin{equation}
\begin{align}
\label{eqn:Pascual_theta}
\nonumber \hat{\theta}_{\text{P}}(x) = & \bigg[45360 (35+17 n)+420 \left(-630-367 n+178 n^2\right) z_{\text{P}} \\
\nonumber& + 3 (n-5) \left(1470-1393 n+430 n^2\right) z_{\text{P}}{}^2 \bigg] \bigg/ \\
\nonumber& \bigg[45360 (35+17 n) + 420 n (-61+178 n) z_{\text{P}} \\
& +5 n \left(3703-919 n+258 n^2\right) z_{\text{P}}{}^2 \bigg].
\end{align}
%\end{equation}
We note that in Pascual's original work, there is a typo in the
numerator in the expression of $\hat{\theta}_{\text{P}}$.  The
number $-376 n$ should be  $-367 n$ instead, and the correct
expression is reproduced here.  The normalised radius
$\hat{\xi}_{\text{P}}$ of the standard LEE as a function of $n$
could be obtained by searching for the first zero of the numerator
of $\hat{\theta}_{\text{P}}(x)$ and inverting the
corresponding value of $x$, i.e., $\hat{\xi}_{\text{P}}$,  from
equation \eqref{eqn:zP}.

On the other hand,
 \citet{Iacono} expressed the  the normalised polytrope function as a
$[1,3]$ Pad\'{e} approximant \citep[see,
e.g.,][]{baker1975essentials} in $x^2$
 using the first five terms of the series
 solution of the LEE:
%\begin{equation}
\begin{align}
\label{eqn:Iacono_theta}
\nonumber \hat{\theta}_{\text{I}}(x) = & \bigg(1- \frac{x^2}{\hat{\xi}_{\text{I}}^2} \bigg) \bigg/ \bigg\{ 1 + x^2 \left(\frac{1}{6} - \frac{1}{\hat{\xi}_{\text{I}}^2} \right) + \frac{x^4}{6} \left(\frac{1}{6} - \frac{n}{20} - \frac{1}{\hat{\xi}_{\text{I}}^2} \right) \\
& + x^6 \left[ \frac{8 n^2 - 47 n +70}{15120} - \frac{10-3 n}{60
\hat{\xi}_{\text{I}}^2} \right] \bigg\},
\end{align}
%\end{equation}
where
\begin{equation}
\label{eqn:Iacono_xi} \hat{\xi}_{\text{I}}(n) = \sqrt{\frac{6
\left(12600-8460 n+1440 n^2\right)}{12600-13490n+4929 n^2-610
n^3}}.
\end{equation}
As a result, the normalised radius of $\hat{\theta}_{\text{I}}(x)$ is
simply equal to $\hat{\xi}_{\text{I}}(n)$, which is expressed as
the square-root of a rational function of $n$ (see equation
\eqref{eqn:Iacono_xi}). It could be easily determined that the
denominator in equation (\ref{eqn:Iacono_xi}) vanishes at $n
\approx 3.05049$, signifying the failure of  the approximant near
and beyond $n \approx 3.05049$.   The approximant is surprisingly
accurate at $n=2$ with the approximate normalised radius
$\hat{\xi}_{\text{I}}(n=2) = \sqrt{360/19} \approx 4.352857$
remarkably close to the numerical value $\hat{\xi}_{\text{exact}}
\approx 4.352875$.

Other approximate expressions of $\hat{\xi}$ similar to that in
equation \eqref{eqn:Iacono_xi} have also been found.
\citet{Buchdahl_n5} found that the normalised radius $\hat{\xi}$ diverges
like a simple pole from the left of $n=5$.  The
idea is to equate two expressions of the potential energy of
polytropes, one involving the normalised polytrope function $\hat{\theta}$
and another involving macroscopic quantities the physical mass $M$,
the physical radius $R$, and polytropic index $n$ only.
Denote the potential energy by $U$, \citet{Buchdahl_n5} showed that:
\begin{equation}
\label{eqn:Buchdahl_proof}
U = - \frac{3 G M^2}{(5 - n) R} = \frac{G M^2}{a (\hat{\xi}^2 \hat{\theta}'(\hat{\xi}))^2} \int_{0}^{\hat{\xi}} x^3 \hat{\theta}(x) \hat{\theta}'(x) dx.
\end{equation}
Substituting the analytical solution of the polytrope function at
$n=5$ into equation (\ref{eqn:Buchdahl_proof}),
\citet{Buchdahl_n5} arrived at equation \eqref{eqn:n5}. By further
interpolating with the analytically exact values of $\hat{\xi}$ at
$n=0$ and $n=1$, \citet{Buchdahl_n5}  proposed a [1,2] Pad\'{e}
approximant \citep[see, e.g.,][]{baker1975essentials} in $n$ for
$\hat{\xi}$:
\begin{equation}
\label{eqn:Buchdahl_xi} \hat{\xi}_{\text{Buch}}(n) =
\frac{12.2474487139 (1 -0.127597320123 n)}{(5 - n) (1
-0.149738026390 n)},
\end{equation}
This simple formula is more accurate than the
approximations derived from the series solution of the LEE for
values of $n>2$, because it does not explicitly depend on the
variation of the polytrope function with the polytropic index $n$,
while the series solution converges slowly near
the surface, and even diverges in the
interior of the polytrope for $n >1.9121$ \citep{Hunter}.

Following \citet{Buchdahl_n5}, \citet{Iacono}
proposed a [2,2] Pad\'{e} approximant \citep[see,
e.g.,][]{baker1975essentials} in $n$ for $\hat{\xi}$:
\begin{equation}
\label{eqn:Iacono_fit_xi} \hat{\xi}_{\text{fit}}(n) =
\frac{12.2378 - 1.2249 n + 0.0187 n^2}{ (5 - n) (1 -0.1223 n)}.
\end{equation}
However, except for the factor $(5-n)$, the coefficients in the
above expression have been determined with nonlinear least-square
fitting to numerical data resulting from Runge-Kutta integrations
for the range where $0 \le n < 5$. Overall, the least-square
fitted approximant is more accurate than that by
\citet{Buchdahl_n5} for the extra free parameter in the numerator,
except near the points $n=0$, $1$ and $5$ where the LEE admits a 
closed form solution for the analytical interpolation.
\citet{Iacono} proposed to replace $\hat{\xi}_{\text{I}}$ by $\hat{\xi}_{\text{fit}}$
in equation (\ref{eqn:Iacono_theta}) to yield an approximation, which is more accurate
near the surface, at a slight compensation of accuracy near the origin, and
we denote such an approximation by $\hat{\theta}_{\text{fit}}$.

\section{List of Useful Formulae} \label{sec:formulae_for_lazy_physicists}
For ease of reference, in the following we quote the global
approximate expressions of the normalised radius $\hat{\xi}$, the
mass $m$, and the normalised polytrope function $\hat{\theta}(x)
$.

\subsection{Radius and normalised radius}
For a polytrope of central density $\rho_c$ under the polytropic
equation of state $P(r)=K[\rho(r)]^{1+1/n}$, its radius is equal
to:
\begin{equation}
R = \sqrt{\frac{K (n+1)}{4 \pi G}} \rho_{c}^{(1-n)/(2n)} \hat{\xi}(n) = a \hat{\xi}(n), \\
\end{equation}
where $\hat{\xi}$ is the first zero of the the normalised
polytrope function, approximately given by the global approximant:
\begin{equation}
\hat{\xi}_{\text{g}}(n) = \frac{ \pi }{1 + a_{0} n^{8}(n-1)^{8}}
\bigg[ \frac{a_{1} + a_{2} n + a_{3} n^2 + a_{4} n^3}{ \sqrt{5-n}
\big(1 + a_{5} n +  a_{6} n^2 \big)} \bigg]^{(n-1)/2},
\end{equation}
with the constants $a_{i}$ for $i=0,1,2,\ldots,6$ determined by
the proposed SDEM as follows:
\begin{equation*}
\begin{split}
& a_{0} \approx 1.5996644405401317 \times 10^{-17}, \\
& a_{1} \approx 3.678184391977817, \\
& a_{2} \approx  -0.12127837785202653, \\
& a_{3} \approx -0.0820898766826553, \\
& a_{4} \approx 0.0030327766768460046, \\
& a_{5} \approx 0.00858273787249898, \\
& a_{6} \approx -0.018845815183087977.
\end{split}
\end{equation*}

The error of $\hat{\xi}_{\text{g}}(n)$ is within $8.1 \times
10^{-7}$ per cent near the incompressible limit for $n$ in
$[0,1]$, and $1$ per cent throughout the physical range of finite
polytropes where $n$ is in $[0,5)$.  The details could be could in
Figure \ref{fig:first_zero}.

\subsection{Mass}
The physical mass $M$ of the same polytrope is given by:
\begin{equation}
M = \int_{0}^{R} \rho(r) 4 \pi r^2 dr = M = \left[\frac{(n+1)K}{4\pi G}\right]^{3/2} \rho_{c}^{(3-n)/(2n)} m(n),
\end{equation}
where $m$ is the dimensionless mass approximately equal to
$m_{\text{g}}(n)$:
\begin{align}
m_{\text{g}}(n) = & b_{0} (5-n)^{(15-3n)/4} n^8 (n-1)^8 \nonumber \\
& + 4 \pi \bigg[ \frac{a_{1} + a_{2} n + a_{3} n^2 + a_{4} n^3}{ \sqrt{5-n} \big(1 + a_{5} n +  a_{6} n^2 \big)} \bigg]^{3(n-1)/2} \nonumber \\
 & \times (5-n)^3 \frac{b_{1} + n b_{2} + n^2 b_{3} + n^3 b_{4}}{1 + n b_{5} + n^2 b_{6}},
\end{align}
where the constants $a_{i}$ for $i=0,1,2,...,6$ are given above
and the constants $b_{i}$ for $i=0,1,2,...,6$ are determined by
the proposed SDEM:
\begin{equation*}
\begin{split}
& b_{0} \approx -3.420867516502784\times 10^{-10}, \\
& b_{1} \approx 0.08268340448079952, \\
& b_{2} \approx  0.0570923774427696, \\
& b_{3} \approx -0.0021371524111317, \\
& b_{4} \approx -0.000863277094516044, \\
& b_{5} \approx 1.370866096910041, \\
& b_{6} \approx 0.415498502167336.
\end{split}
\end{equation*}

The error is within $8.5 \times 10^{-5}$ per cent for $n$ in
$[0,1]$, and $2$ per cent throughout the physical range of finite
polytropes where $n$ is in $[0,5)$. The details could be could in
Figure \ref{fig:mass}.

\subsection{Density and normalised polytrope function}
For a polytrope of central density $\rho_{c}$, the normalised
polytrope function $\hat{\theta}(x)$ is related to the density
profile of the polytrope $\rho(r) = \rho_{c}
[\hat{\theta}(r/a)]^n$. Using the scaled variables $z_{\text{g}} =
(\pi r) / (a \hat{\xi}_{\text{g}})$ and $z_{\text{P}} = 6 [ ( 1 +
x^2/3 )^{1/2} -1 ]$, the normalised polytrope function is
approximated piecewise using $\hat{\theta}(x) \approx
\hat{\theta}_{\text{g}}(x)$, where:
\begin{equation}
\label{eqn:theta_global}
\hat{\theta}_{\text{g}} (x) = \begin{cases}
\hat{\theta}_{\text{tp}}(x), &\text{for} \qquad 0 \leq n \leq 2, \\
\hat{\theta}_{\text{P}}(x), &\text{for} \qquad 2 < n < 5,
\end{cases}
\end{equation}
with the functions $\hat{\theta}_{\text{P}}(x)$ and
$\hat{\theta}_{\text{tp}}(x)$ given by:
\begin{align}
\nonumber \hat{\theta}_{\text{P}}(x) = & \bigg[45360 (35+17 n)+420 \left(-630-367 n+178 n^2\right) z_{\text{P}} \\
\nonumber& + 3 (n-5) \left(1470-1393 n+430 n^2\right) z_{\text{P}}{}^2 \bigg] \bigg/ \\
\nonumber& \bigg[45360 (35+17 n) + 420 n (-61+178 n) z_{\text{P}} \\
& +5 n \left(3703-919 n+258 n^2\right) z_{\text{P}}{}^2 \bigg],
\end{align}
\begin{equation}
\begin{split}
\hat{\theta}_{\text{tp}}(x) = & \Theta_0^{(0)}(z_{\text{g}})+n \Theta_0^{(1)}(z_{\text{g}})+n^2 \Theta_0^{(2)}(z_{\text{g}}) +n^3 \bigg[-4 \Theta_0^{(0)}(z_{\text{g}}) \\
& -3 \Theta_0^{(1)}(z_{\text{g}}) -2 \Theta_0^{(2)}(z_{\text{g}}) + 4 \Theta_1^{(0)}(z_{\text{g}})-\Theta_1^{(1)}(z_{\text{g}})\bigg] \\
& +n^4 \bigg[3 \Theta_0^{(0)}(z_{\text{g}}) +2 \Theta_0^{(1)}(z_{\text{g}})+\Theta_0^{(2)}(z_{\text{g}})\\
& -3
\Theta_1^{(0)}(z_{\text{g}})+\Theta_1^{(1)}(z_{\text{g}})\bigg].
\end{split}
\end{equation}
In the above the  $\Theta$ functions are obtained perturbatively
from the SDEM:
\begin{equation}
\Theta_{0}^{(0)} (z) = 1 - \frac{z^2}{\pi^2},
\end{equation}
\begin{align}
\nonumber \Theta_{0}^{(1)}(z) = & -4 + \frac{4 z^2}{\pi^2} \left(1 - \ln 2 \right)
+\left(3-\frac{2 \pi }{z}-\frac{z^2}{\pi ^2}\right) \ln \left(1-\frac{z}{\pi}\right)\\
& +\left(3+\frac{2 \pi }{z}-\frac{z^2}{\pi ^2}\right) \ln
\left(1+\frac{z}{\pi }\right),
\end{align}
\begin{align}
\Theta_{0}^{(2)}(z) = & 40+\frac{7 \pi ^2}{3}+8 \ln 2-14 \ln ^2 2 \nonumber \\
& + \left[\frac{7 \pi^2}{3}- 40 +32 \ln 2 -8 \ln ^2 2 \right] \frac{z^2}{\pi^2} \nonumber \\
& + \left[-23+\frac{20 \pi }{z}+\frac{3 z^2}{\pi ^2}+ \big(14 -\frac{10 \pi}{z}-\frac{4 z^2}{\pi ^2}\big)\ln 2 \right]\nonumber \\
& \quad \times \ln \left(1-\frac{z}{\pi}\right) + \left(\frac{3}{2}-\frac{\pi }{z}-\frac{z^2}{2 \pi ^2}\right) \ln ^2\left(1-\frac{z}{\pi }\right) \nonumber \\
& + \left[-23-\frac{20 \pi }{z}+\frac{3 z^2}{\pi ^2}+\big( 14+\frac{10 \pi}{z}-\frac{4 z^2}{\pi ^2}\big) \ln 2 \right] \nonumber \\
& \quad \times \ln \left(1+\frac{z}{\pi}\right) + \left(\frac{3}{2}+\frac{\pi }{z}-\frac{z^2}{2 \pi^2}\right) \ln ^2\left(1+\frac{z}{\pi }\right) \nonumber \\
& + \left(1-\frac{z^2}{\pi ^2}\right) \ln \left(1-\frac{z}{\pi }\right) \ln \left(1+\frac{z}{\pi }\right)\nonumber \\
& + \left(\frac{14 \pi }{z}-14\right) \text{Li}_2\left(\frac{\pi -z}{2 \pi }\right)\nonumber \\
& + \left(-\frac{14 \pi }{z}-14\right) \text{Li}_2\left(\frac{\pi
+z}{2 \pi}\right),
\end{align}
\begin{equation}
\Theta_{1}^{(0)}(z) = \frac{\sin z}{z},
\end{equation}
\begin{align}
\Theta_{1}^{(1)}(z) & = \frac{\sin z}{z} \bigg[ 1-\frac{\ln(2\pi)}{2}-\frac{\text{Si}(2 \pi )}
{4 \pi } - \frac{\text{Cin}(2 z)}{4} \nonumber \\
& +\frac{\ln z}{2}-\frac{1}{2} \ln(\sin z) \bigg] + \frac{\cos z}{z} \bigg[ \frac{1}{2} z \ln (2 \pi) \nonumber \\
& -\frac{1}{2} z \ln z+\frac{z \text{Si}(2 \pi )}{4 \pi
}-\frac{\text{Si}(2 z)}{4} + \frac{1}{2} \int_{0}^{z} \ln \sin t
dt \bigg].
\end{align}

The root mean square error of the approximant
$\hat{\theta}_{\text{g}} (x)$ is within $8.4 \times 10^{-6}$ for
$n$ in $[0,1]$, and $6.1 \times 10^{-3}$ for $n$ in $[0,5)$.  For
better illustrations, readers may refer to Figures
\ref{fig:n15_theta} and \ref{fig:theta_global_error}.
%%%%%%%%%%%%%%%%%%%%%%%%%%%%%%%%%%%%%%%%%%%%%%%%%%

% Don't change these lines
\bsp    % typesetting comment
\label{lastpage}
\end{document}